\documentclass[floats,floatfix,amssymb,prd,twocolumn,superscriptaddress,nofootinbib]{revtex4-1}
\usepackage{cancel}
\usepackage{slashed}
\usepackage{subcaption,tensor}
\usepackage{ragged2e}
\DeclareCaptionJustification{justified}{\justifying}
\makeatletter
\newcommand{\subsetsim}{\mathrel{\mathpalette\subset@sim\relax}}
\newcommand{\subset@sim}[2]{%
  \vtop{\offinterlineskip\m@th
    \ialign{\hfil##\cr
      $#1\subset$\cr\noalign{\kern0.5pt}\scalebox{0.9}{$#1\sim$}\cr
    }%
  }%
}
\makeatother
 
\usepackage{amssymb,amsmath,verbatim,mathtools,needspace,enumitem,etoolbox,graphicx,physics,microtype,afterpage,bm}
\usepackage[dvipsnames, usenames]{xcolor}
\definecolor{linkcolor}{rgb}{0.0,0.3,0.5}
\usepackage{booktabs}
\definecolor{brilliantlavender}{rgb}{0.96, 0.73, 1.0}
\definecolor{oucrimsonred}{rgb}{0.6, 0.0, 0.0}
\definecolor{persianblue}{rgb}{0.11, 0.22, 0.73}
\definecolor{forestgreen}{rgb}{0.13,0.35,0.13}
\definecolor{firebrick}{rgb}{0.7, 0.13, 0.13}
\usepackage[unicode, 
colorlinks=true, 
linkcolor=oucrimsonred, 
citecolor=oucrimsonred, 
filecolor=oucrimsonred,
urlcolor=oucrimsonred, 
pdfusetitle]{hyperref}

\usepackage[all]{hypcap}
\usepackage[T1]{fontenc}
\usepackage[utf8]{inputenc}
\usepackage{tabularx,comment}
\usepackage{adjustbox}
\usepackage{float}
\usepackage{ulem}
\usepackage{xfrac}
\usepackage{orcidlink}
\usepackage{bbm}

\tikzset{->-/.style={decoration={
  markings,
  mark=at position #1 with {\arrow[scale=1.4]{Latex}}},postaction={decorate}}}
  \tikzset{-<-/.style={decoration={
  markings,
  mark=at position #1 with {\arrowreversed[scale=1.4]{Latex}}},postaction={decorate}}} 
\tikzset{--<--/.style={decoration={
  markings,
  mark=at position #1 with {\arrowreversed[scale=1.2,rotate=18]{Latex}}},postaction={decorate}}} 

\renewcommand{\arraystretch}{1.4}

\definecolor{azure}{rgb}{0.0, 0.5, 1.0}
\definecolor{deepfuchsia}{rgb}{0.76, 0.33, 0.76}
\definecolor{VioletRed4}{rgb}{0.55, 0.13, .32}
 
\definecolor{harvardcrimson}{rgb}{0.79, 0.0, 0.09}
\definecolor{oceanboatblue}{rgb}{0.0, 0.47, 0.75}
\definecolor{persianblue}{rgb}{0.11, 0.22, 0.73}
\definecolor{egyptianblue}{rgb}{0.06, 0.2, 0.65}
\definecolor{navyblue}{rgb}{0.0, 0.0, 0.5}

\definecolor{verdechiaro}{rgb}{0.6,1,0.6}
\definecolor{giallochiaro}{rgb}{1,1,0.6}
\definecolor{bluscuro}{rgb}{0.15, 0.2, 0.9}
\definecolor{verdes}{rgb}{0.1, 0.5, 0.1}%
\definecolor{tangerineyellow}{rgb}{1.0, 0.8, 0.0}

\usepackage{multirow}
\usepackage{pifont}
\usepackage{fontawesome}
\usepackage{lmodern}

\usepackage{multirow}

\allowdisplaybreaks
\usepackage{tikz}
\usepackage{tikz-feynman}
\usetikzlibrary{tikzmark}
\usepackage{tcolorbox}
\usepackage{pifont}
\usepackage{color}
\usepackage{framed}

\definecolor{rossos}{cmyk}{0,1,1,0.55}
\definecolor{bluscuro}{rgb}{0.15, 0.2, .85}
\definecolor{bluchiaro}{cmyk}{1,.3,0.,0.1}
\definecolor{ForestGreen}{rgb}{0.13, 0.55, 0.13}

\newtcolorbox{mybox}{colback=mycolor!5!white,colframe=azure!75!black}

\def\f{\frac}

\def\nn{\nonumber}

\def\bea{\begin{eqnarray}}
\def\eea{\end{eqnarray}}

\newcommand{\bs}{\begin{subequations}}
\newcommand{\es}{\end{subequations}}

\newcommand{\be}{\begin{equation}}
\newcommand{\ee}{\end{equation}}

\def\lsim{\mathrel{\rlap{\lower4pt\hbox{\hskip0.5pt$\sim$}}
    \raise1pt\hbox{$<$}}}         
\def\gsim{\mathrel{\rlap{\lower4pt\hbox{\hskip0.5pt$\sim$}}
    \raise1pt\hbox{$>$}}}         

\makeatletter
\def\l@subsubsection#1#2{}
\makeatother

\newcommand{\sapienza}{Dipartimento di Fisica, Sapienza Università 
	di Roma, Piazzale Aldo Moro 5, 00185, Roma, Italy}
\newcommand{\infn}{INFN Roma\,1, Piazzale Aldo Moro 2, 00185, Roma, Italy}

\begin{document}

\title{
Gravitational instantons and the quality problem of the QCD axion:\\
Facts, speculations, and statements in between
}

\author{Pier Giuseppe Catinari}
\email{piergiuseppe.catinari@uniroma1.it}
\affiliation{\sapienza}
\affiliation{\infn}

\author{Alfredo Urbano\orcidlink{0000-0002-0488-3256}}
\email{alfredo.urbano@uniroma1.it}
\affiliation{\sapienza}
\affiliation{\infn}

\date{\today}

\begin{abstract}
In this work, we critically reanalyze the explicit breaking of the Peccei-Quinn global symmetry---and the corresponding corrections to the QCD axion potential---induced by gravity. Specifically, we examine the role of gravitational instantons, which are non-perturbative, finite-action solutions to the Euclidean Einstein equations. These instantons represent topologically nontrivial configurations of spacetime and are analogous to instantons in gauge theory. The amount of symmetry breaking induced by gravitational instantons can be computed in a controlled way within the framework of semi-classical gravity, using 't Hooft operators, in full analogy to the computation of the axion potential arising from QCD small instanton effects. Contrary to previous results in the literature, we find that the effects of gravitational instantons are extremely small and therefore do not give rise to a significant quality problem for the axion solution to the strong CP problem, both within the Standard Model and in beyond-the-Standard-Model scenarios that involve multiple copies of the Standard Model. In conclusion, we argue that, assuming the ultraviolet completion of gravity is weakly coupled, the axion solution to the strong CP problem remains free from any quality issues due to gravity. Along the way, we derive the effective Lagrangian of the QCD axion, including its gravitational coupling.
\end{abstract}

\maketitle

{
  \hypersetup{linkcolor=black}
}

\normalem


\section{Introduction}\label{sec:Intro}

The strong CP problem in quantum chromodynamics (QCD)  arises from the fact that QCD allows for a CP-violating term in its Lagrangian, proportional to $\theta G^a_{\mu\nu}\tilde{G}^{a,\mu\nu}$, where $\theta$ is a parameter, and $G^a_{\mu\nu}\tilde{G}^{a,\mu\nu}$  represents the topological term related to the gluon field strength tensor $G^a_{\mu\nu}$ and its Hodge dual $\tilde{G}^{a}_{\mu\nu}$. 
The dual is defined as
$\tilde{G}^{a}_{\mu\nu} = 
\varepsilon_{\mu\nu\rho\sigma}G^{a,\rho\sigma}/2$,  where $\varepsilon_{\mu\nu\rho\sigma}$ is the totally antisymmetric Levi-Civita symbol, with $\varepsilon_{0123} = +1$.
Experimental results, however, show no evidence of CP violation in strong interactions, implying that $\theta$ must be extremely small, which is puzzling given the lack of any fundamental reason for it to be zero.

The Peccei-Quinn (PQ) mechanism addressing the strong CP problem faces a significant challenge: it depends on a global $U(1)_{\textrm{PQ}}$ symmetry spontaneously broken. This symmetry is introduced to dynamically solve the strong CP problem by promoting the CP-violating $\theta$-parameter in QCD to a dynamical field, the axion, which adjusts itself to cancel the CP violation. 
Although this symmetry is broken at low energies by the QCD anomaly, it must remain an exceptionally precise symmetry at high-energy scales. This issue is commonly referred to as the PQ quality problem. 
This fact seems to contradict the modern effective field theory understanding that global symmetries are not fundamental but rather emerge as accidental symmetries, namely symmetries of the renormalizable Lagrangian whose violation occurs through irrelevant operators.

This conceptual problem is further exacerbated by the widely held belief that gravity explicitly breaks  global symmetries. If true, this would imply that any global symmetry present in a low-energy effective theory would be violated by gravitational interactions, potentially undermining the robustness of such symmetries at high-energy scales.

In the aforementioned context, the quality problem can be readily quantified as follows. If Planck-scale physics is the sole source of PQ-violating effects, then at energies below the Planck scale, the effect of gravitational interactions is encoded by the effective operator in the axion Lagrangian\,\cite{Kamionkowski:1992mf}

\begin{equation}
    \frac{1}{\sqrt{-g}}\mathcal{O}_g(x)=\sum_{n\geq5}\frac{\lambda^2}{M_{\rm{Pl}}^{n-4}}\Phi^n(x)+\rm{h.c.}+c,
\end{equation}
where $\Phi(x)=f_ae^{i a/f_a}$, 
$f_a$ is the scale at which the $U(1)_{\textrm{PQ}}$ symmetry gets spontaneously broken, $\lambda=|\lambda |\exp(i\delta)$ is a dimensionless complex coupling and $c$ is a constant chosen so that the minimum of the potential is zero. The effective operator $\mathcal{O}_g$ then induces the axion effective potential
\begin{align}\label{eq:gravitational effective potential}
V_g(a)&=2|\lambda |^2 M_{\rm{Pl}}^4\times\nn\\
&\times\sum_{n\geq5}\left(\frac{f_a}{M_{\rm{Pl}}}\right)^{n}\left[1-\cos\left(n\frac{a}{f_a}+2\delta\right)\right]\,,
\end{align}
therefore, for $O(1)$ values of the phase $\delta$ and $|\lambda|$, taking $f_a = 10^{12}~\rm{GeV}$, the $U(1)_{\textrm{PQ}}$ symmetry must be protected from explicit breaking effects caused by operators with dimensions up to $d \sim 14$, suggesting a highly non-trivial structure in the PQ dynamics. We refer to this situation, where explicit breaking effects are introduced through irrelevant operators whose scaling is determined by naive dimensional analysis, as a case of perturbative breaking.\\This specification is necessary, as it is by no means certain that gravity breaks global symmetries as described above. In fact, gravitational interactions are CP-conserving perturbatively, and the only conceivable source of CP violation is nonperturbative quantum gravity dynamics. Multiple indications indeed suggest that explicit breakings due to gravity have a nonperturbative origin.

\subsection{Symmetry breaking due to gravitational and stringy effects}\label{sec:StringyEffects}

Arguments in black hole physics, often referred to as folklore theorems \cite{Banks:2010zn,Susskind:1995da}, suggest that black holes with continuous global charge would result in an infinite number of macroscopically indistinguishable Planck-scale remnants, in clear contradiction to black hole thermodynamics \cite{Bousso:1999cb,Bousso:2002ju}.
Secondly, in the context of string theory, it can be shown that explicit breaking effects of continuous global symmetries are always due to non-perturbative effects, with candidates including worldsheet instantons, brane instantons and gauge instantons\,\cite{Banks:1996ea,Svrcek:2006yi}.
The inability to preserve a global symmetry in string theory can be seen as a consequence of the fact that the effective field theories associated with classical vacua of superstring theories (regardless of whether they are spacetime supersymmetric) lack continuous global symmetries—all continuous symmetries are gauged\,\cite{Banks:1988yz}.
Alternatively, in the context of the AdS/CFT correspondence, it can be demonstrated that any global symmetry, whether discrete or continuous, in a bulk quantum gravity theory  would be inconsistent with entanglement wedge reconstruction on the CFT side of the correspondence\,\cite{Harlow:2018tng}.

We now briefly discuss some key aspects of a string-theoretic setup, as certain details will prove important in the continuation of our work. 
In the perturbative formulation of string theory, the string mass scale 
$M_s$ and the string coupling $g_s$ are fundamental parameters that govern both the physics of the strings themselves and their interactions. 
The string mass scale $M_s$ 
is a measure of the energy at which stringy effects become significant.  
The string coupling $g_s$ controls the strength of interactions between strings. In the weak-coupling regime, where $g_s \ll 1$, the interactions between strings are rare, and the perturbative expansion in powers of $g_s$ is well-defined. 
The Planck mass of gravity in four-dimensional spacetime is related to $M_s$, $g_s$, and the six-dimensional 
volume of the compactified extra dimensions $V$. When six of the original ten dimensions of string theory are compactified, the four-dimensional Planck mass is related to the string scale by\,\cite{Antoniadis:1998ig}
\begin{align}
M_{\textrm{Pl}}^2 \approx  \frac{(V M_s^6) M_s^2}{g_s^2}\,.    
\end{align}
In addition to the string mass scale, another important energy scale in compactified string theories is the Kaluza-Klein (KK) mass scale. This scale arises from the compactification of extra dimensions, where the momentum modes of particles propagating in the compact dimensions are quantized due to the compact geometry. The KK mass scale is inversely related to the size $R$ of the compactified dimensions and is typically given by 
$M_{\textrm{KK}} \approx  R^{-1} \approx  V^{-1/6}$ (under the assumption that the 
compactified dimensions have a common radius $R$). We thus find
\begin{align}
g_s M_{\textrm{Pl}} \approx 
\left(
\frac{M_s}{M_{\textrm{KK}}}
\right)^3 M_s\,.
\end{align}
It is important to emphasize that the two mass scales $M_s$ (related to the fundamental string tension and controlling the onset of stringy effects---i.e., the energy scale at which the extended nature of strings becomes significant and point-particle approximations break down) and $M_{\textrm{KK}}$ (determined by the size $R$ of the compactified extra dimensions) are generally independent parameters in a string compactification,\footnote{For example, in large extra dimension scenarios, $M_{\textrm{KK}}$ can be much smaller than $M_s$, allowing KK modes to become observable at lower energies.} although  they can become comparable under certain conditions. 
If we consider the case $M_s \approx M_{\textrm{KK}}$, the string mass scale, the string coupling and the four-dimensional Planck mass are related by  
\begin{align}
    g_s M_{\textrm{Pl}}\approx M_s\,,\label{eq:PlanckMassCoupling}
\end{align}
with 
$M_s \ll M_{\textrm{Pl}}$ in the weakly coupled regime.
Under these circumstances, and based solely on dimensional analysis, the breaking effects due to non-perturbative stringy instantons are expected to contribute to the axion potential as in \eqref{eq:gravitational effective potential}, now with $|\lambda|^2=g_s^2e^{-C/g_s}$
where 
$C=O(1)$ is an order 1 constant.

The key aspect is the exponential suppression given by the factor $e^{-{S_{\textrm{E}}}} = e^{-C/g_s}$, where $S_{\textrm{E}}$ is the Euclidean action of the string instanton generating the axion potential\,\cite{Shenker1991}. 
Consequently, if string theory is weakly coupled, i.e., if $g_s \ll 1$, all non-perturbative effects, and in particular the explicit symmetry-breaking effects of global symmetries, are exponentially suppressed. 
In this regard, it is worth emphasizing that there are strong semi-quantitative indications suggesting that quantum gravity is indeed ultraviolet (UV) completed at weak coupling.
{\it i)} Imagine $N$ species of quantum fields, with masses at the
scale $M$, coupled to gravity. 
In this case, 
both perturbative renormalization arguments\,\cite{Larsen:1995ax,Calmet:2008tn} and non-perturbative black hole considerations\,\cite{Dvali:2007hz,Dvali:2007wp} suggest that in the presence of $N$-species the effective contribution to the Planck mass is parametrically given by  
$(N/16\pi^2)M^2 \approx M_{\textrm{Pl}}^2$ which would imply 
$g_s \approx 4\pi/\sqrt{N}$. 
Consequently, the presence of a large number of particle species, $N\gg 1$, lowers the gravitational cutoff parametrically below the Planck mass.
{\it ii)} In the Standard Model, flavor symmetries are approximate global symmetries explicitly broken by the Yukawa couplings. From the perspective of the UV theory, the smallness of certain Yukawa couplings appears to be compatible with the exponential suppression associated with the presence of weak coupling.
{\it iii)} 
Inflation is a theoretical phase of rapid exponential expansion in the early universe, proposed to resolve issues like the horizon, flatness, and monopole problems. It is driven by the potential energy of a scalar field, the inflaton, which causes space to expand dramatically over a very short time. This expansion amplifies quantum fluctuations, stretching them to macroscopic scales and seeding the anisotropies observed in the cosmic microwave background (CMB). The energy scale of inflation is characterized by the Hubble parameter $H$, typically around $H= O(10^{13})$ GeV, though it varies with different inflationary models. The amplitude of the scalar power spectrum, which describes the density fluctuations in the early universe, is related to $H$; specifically, the larger $H$, the larger the predicted amplitude of these primordial fluctuations. The observed amplitude of the scalar power spectrum, $A_s \simeq 2.1 \times 10^{-9}$, provides constraints on the inflationary energy scale and potential. 
In string theory, the inflaton field can emerge from various configurations, such as the distance between branes or compactification moduli\,\cite{Baumann:2014nda}.  
Most
models of inflation in string theory are formulated under the condition 
$H\ll M_s \approx M_{\textrm{KK}} \ll M_{\textrm{Pl}}$, with the 
inflaton mass, $m_{\phi}$, parametrically given by
$m_{\phi} \approx \sqrt{\eta_V} H$, where 
$\eta_V = M_{\textrm{Pl}}^2 V^{\prime\prime}(\phi)/V(\phi)$, with $V(\phi)$ being the inflaton potential. 
For slow-roll inflation to occur, the potential must be relatively flat, which means that the inflaton mass is typically smaller than or comparable to the Hubble scale, $\eta_V \ll 1$. 
Under these assumptions, the inflaton is the only light field, while the others have masses far above the Hubble scale; the heavy fields can then be integrated out, leaving a model of single-field inflation. Describing the dynamics of the inflaton within the context of an effective theory featuring only a single mass scale, $M_s$, and a fundamental coupling, $g_s$, leads to the result that the amplitude of the scalar power spectrum is intrinsically tied to the coupling $g_s^2$.
This is a consequence of dimensional analysis. In the slow-roll approximation, we write 
$A_s = H^2/8\pi^2\epsilon \bar{M}_{\textrm{Pl}}^2$, where $\epsilon$ is the first slow-roll parameter and $\bar{M}_{\textrm{Pl}}$ the reduced Planck mass. 
The ratio $H^2/\bar{M}_{\textrm{Pl}}^2$ has dimensions of a coupling squared, and the only possibility---given the assumptions under which we are working---is that $A_s \propto g_s^2$.
Consequently, the measurement of $A_s$ implies that the string coupling is in the weak regime. 

Throughout this work, we will assume that the UV completion of gravity is a string theory in the perturbative regime of weak coupling.

\subsection{Symmetry breaking due to Euclidean wormholes}\label{sec:Vermi}

Finally, it has been conjectured that global symmetries are violated by Euclidean wormholes in the quantum gravity path integral\,\cite{Giddings:1987cg,Abbott:1989jw,Kallosh:1995hi,Alonso:2017avz,Hsin:2020mfa}.   
If this claim were true, the breaking effect due to wormholes could pose a threat to the axion solution of the strong CP problem; 
it is indeed known that the scalar action describing the spontaneous breaking of a $U(1)$ symmetry admits wormhole solutions (dubbed Euclidean axion wormholes), whose explicit symmetry-breaking effects become particularly significant when the radial degree of freedom of the complex scalar field is treated as dynamic\,
\cite{Kallosh:1995hi,Alonso:2017avz,Alvey:2020nyh}.
However, there are some very fundamental conceptual issues with Euclidean wormholes, cf. ref.\,\cite{Hebecker:2018ofv} for a recent review. Specifically, it is not yet clear whether Euclidean wormholes actually contribute to the path integral.
It is indeed crucial to ensure that the wormhole solutions are stable to small perturbations of their topology, an analysis which so far has not led to definitive conclusions. 
See, for example, the discussion in ref.\,\cite{Hertog:2018kbz}, where it is concluded that wormholes are not relevant saddle points of the functional integral in quantum gravity.

Throughout this work, we will not take into account the presence of Euclidean wormholes, assuming that they do not contribute to the quantum gravity path integral.

\subsection{Gravitational instantons}\label{sec:GI}

Given these premises, the perspective we choose to consider in this work is as follows.
Let us first return to the case of QCD. 
Yang-Mills instanton solutions---non-perturbative solutions to the Yang-Mills equations that describe tunneling between different vacua of QCD with different topological charges\,\cite{JackiwRebbi,CallanDashenGross}---are essential for the axion solution to the strong CP problem since they provide the non-perturbative dynamics that generate the axion potential, allowing the axion field to dynamically cancel the $\theta$-term and restore CP symmetry in QCD. 

Remarkably, Einstein's theory of gravitation, in its Euclidean formulation, admits non-perturbative solutions, dubbed gravitational instantons, that 
strongly resemble the Yang-Mills instantons\,\cite{Hawking:1976jb,Gibbons:1979xm}. 
Consequently, it is reasonable to expect that these gravitational instantons are the most likely candidates to play a significant role when considering gravitational effects in the context of the axionic solution to the strong CP problem. 
These solutions, in fact, persist even under the assumption that string theory is weakly coupled (cf. section\,\ref{sec:StringyEffects}) or when neglecting the effects of wormholes (cf. section\,\ref{sec:Vermi}), and, as we will demonstrate, provide a calculable contribution to the axion potential in the controlled setup of semi-classical gravity.

Let us try to clarify more precisely what a gravitational instanton actually is (cf. ref.\,\cite{Eguchi:1980jx} for a comprehensive discussion).
In the fields of mathematical physics and differential geometry, a gravitational instanton refers to a non-singular, 
 geodesically  complete, four-dimensional Riemannian manifold that solves the vacuum Einstein equations $R_{\mu\nu} = 0$ (without a cosmological constant, also known as  Ricci-flat metrics). 
These manifolds are called gravitational instantons due to their role as analogues to instantons from Yang–Mills theory, but in the realm of quantum gravity. Similar to self-dual Yang–Mills instantons, gravitational instantons are typically expected to behave like four-dimensional Euclidean space at large distances and possess a self-dual Riemann tensor.  
A self-dual Riemann tensor verifies 
the condition $R_{\mu\nu\rho\sigma} = \tilde{R}_{\mu\nu\rho\sigma}$ 
with $\tilde{R}_{\mu\nu\rho\sigma} = 
\epsilon_{\mu\nu}^{~~\,\alpha\beta}R_{\alpha\beta\rho\sigma}/2$ and 
$\epsilon_{\mu\nu\alpha\beta} = \sqrt{g}\varepsilon_{\mu\nu\alpha\beta}$ the totally antisymmetric Levi-Civita tensor (with $\epsilon^{\mu\nu\alpha\beta} = \varepsilon^{\mu\nu\alpha\beta}/\sqrt{g}$).
We remark that this condition automatically guarantees the vanishing of the Ricci tensor, $R_{\mu\nu} = 0$.
In addition, we focus on solutions that have the interesting property of approaching a flat metric at infinity. 
Since the Yang-Mills instanton potential approaches a pure gauge configuration at infinity, this class of Einstein solutions closely resembles the Yang-Mills case. 
Ricci-flat, self-dual metrics approaching a flat metric at infinity exhibit several notable characteristics. Firstly, they typically  possess zero action, which highlights their importance in the path integral approach to quantum gravity.\footnote{
 The Euclidean Einstein-Hilbert action is given by
 \begin{align}
S = \underbrace{\frac{1}{16\pi G_N}\int d^4 x 
\sqrt{g}\,R}_{\equiv\,S_{\textrm{bulk}}} + S_{\textrm{boundary}}\,,
 \end{align}
 where $R$ is the Ricci scalar, $g$ is the determinant of the metric, and $S_{\textrm{boundary}}$ represents the boundary contribution given by the Gibbons-Hawking-York boundary term\,\cite{York:1972sj,Gibbons:1976ue}. 
 Ricci-flat gravitational instantons have $R=0$, and the bulk part of the action vanishes. 
 However, it is possible to get a non-vanishing contribution from the boundary term.
} Additionally, as these metrics tend toward flatness and effectively eliminate gravitational interactions at infinity, one can employ standard asymptotic-state techniques to investigate quantum effects. 
On the contrary, self-dual metrics that are regular and have finite action, but are not asymptotically flat, are characterized by a gravitational field that persists throughout spacetime, making it difficult to define the asymptotic plane-wave states necessary for ordinary scattering theory.

There are no non-trivial gravitational instantons that are truly asymptotically Euclidean\,\cite{Schon:1979rg,Witten:1981mf}.\footnote{More precisely, if the manifold has the topology of Euclidean space at infinity, that is, $S^3\times \mathbb{R}$, and approaches the flat Euclidean metric sufficiently fast, then every self-dual solution is isometric to Euclidean space, cf. ref.\,\cite{Gibbons:1979xn}.} However, there are two important classes of gravitational instantons that approach flat space at infinity in a specific manner.
\begin{itemize}
\item[$\circ$] Asymptotically locally Euclidean (ALE) metrics. 
These solutions are called ALE metrics because, despite their local flatness at large distances, their global topology at infinity differs from that of ordinary Euclidean space.
More specifically, ALE metrics are asymptotically flat, but the boundaries are three-spheres with points
identified under the action of some discrete group, namely $(S^3/\Gamma)\times \mathbb{R}$ where $\Gamma$ is a finite group of isometries acting freely on $S^3$.

Notable examples include the Eguchi-Hanson (EH) metric\,\cite{Eguchi:1978gw} and the Gibbons-Hawking multi-center metric in special configurations in which the centers are arranged symmetrically\,\cite{Gibbons:1978tef}.
\item[$\circ$] Asymptotically locally flat (ALF) metrics. 
ALF spaces also approach flat space at infinity, but their asymptotic structure is more complicated. In ALF spaces, the geometry at infinity resembles a flat space with a circle fibration.

Notable examples include the Taub-NUT space\,\cite{Taub:1950ez,Newman:1963yy}, where the metric asymptotically approaches $\mathbb{R}^3\times S^1$, and the Gibbons-Hawking multi-center metric in generic configurations.
\end{itemize}
In general, Ricci-flat ALE metrics---such as the EH metric---exhibit zero action, while Ricci-flat ALF metrics---such as the Taub-NUT metric (cf. ref.\,\cite{Gibbons:1979xm})---possess finite non-zero action due to their more complex asymptotic structure, which introduces a non-zero contribution from the boundary term. 
In a semiclassical approach to quantum gravity, therefore, ALE instantons are often considered more relevant than ALF instantons due to their zero action. 
For this reason, we will focus on ALE gravitational instantons for the remainder of this work (we will provide a more detailed justification in section\,\ref{sec:NonPerturbativeGravity}). Topologically speaking, there are three invariants of utmost importance: the Euler characteristic $\chi$, the Hirzebruch signature $\tau$, and the spin-$\frac{1}{2}$ index $I_{\frac{1}{2}}$.
These three invariants are central in characterizing the properties of smooth manifolds and can be thought of as the gravitational analogs of the Yang-Mills topological invariants\,\cite{Belavin:1976rx,Marciano:1976hp,Eguchi:1976db}.
Gravitational instantons have been classified based on their topological properties in refs.\,\cite{Gibbons:1979xm,Eguchi:1980jx}.
Gravitational instantons with non-zero spin-$\frac{1}{2}$ index are those that support non-trivial solutions to the massless Dirac equation for spin-$\frac{1}{2}$ fermions (called fermionic zero modes) in the background of the gravitational instanton.  
In this regard, there is once again a strong parallelism between gravity and Yang-Mills theories. 
In Yang-Mills theory, fermions couple to gauge fields, and their dynamics are governed by the Dirac equation in the background of the gauge field. For a given instanton solution of the Yang-Mills equations, the Dirac operator in that background may have zero modes-solutions to the Dirac equation with zero eigenvalue. These zero modes are essential because instantons induce processes that can violate chiral symmetry in theories with massless fermions.
Therefore, we are particularly interested in gravitational instantons that possess a non-zero spin-$\frac{1}{2}$ index.
However, most of the known gravitational instantons have a vanishing spin-$\frac{1}{2}$ index\,\cite{Eguchi:1980jx}.
The only relevant exception is the gravitational instanton known as K3. 
The K3 surface is the only compact regular simply-connected manifold without
boundary which admits a nontrivial metric with self-dual curvature. The spin-$\frac{1}{2}$ index is 
$I_{\frac{1}{2}}(\textrm{K3}) = +2$. However, as just mentioned, K3 has no boundaries and does not approach a flat space at infinity.
It is still possible to take K3-type gravitational instantons into account through a procedure that involves cutting out a 4-dimensional ball and gluing K3 into flat spacetime via a wormhole-type throat\,\cite{Hebecker:2019vyf}.  
The downside is that K3 transitions from having a vanishing action to a Planckian action of the form $S_{\textrm{E}} \sim \rho^2 M_{\textrm{Pl}}^2$\,\cite{Hebecker:2019vyf}, where $\rho$ is the typical size of the wormhole throat, and its contribution to the gravitational path integral becomes automatically suppressed. For this reason, we will not further analyze this possibility in this work.\footnote{Moreover, as already recognized in the analysis of ref.\,\cite{Hebecker:2019vyf}, it is unclear whether the stability problem of the wormhole mentioned earlier, if present, could give rise to any pathology in this patched solution as well.}

However, there exists a second, natural possibility: one can seek and construct a combined solution to the Maxwell-Einstein equations in the presence of the gravitational instanton metric. 
Remarkably, introducing a gravitational instanton that also solves the Maxwell equations can lead to significant changes in the spin-$\frac{1}{2}$ index, often resulting in non-vanishing contributions due to the interplay between fermions and the electromagnetic field in the modified spacetime. 
In particular, this is the case for the ALE EH gravitational instanton\,\cite{Eguchi:1978gw}. 
The latter has Euler characteristic $\chi(\textrm{EH}) = 2$, 
Hirzebruch signature $\tau(\textrm{EH}) = -1$, 
spin-$\frac{1}{2}$ index $I_{\frac{1}{2}}(\textrm{EH}) = 0$ and vanishing action. 
The EH gravitational instaton supports a non-zero Maxwell field strenght  
with $1/r^4$ asymptotic behavior. 
It is worth emphasizing that this asymptotic behavior is the same as that of Yang-Mills instantons.\footnote{For instance, the self-dual ALF Taub-NUT metric supports a non-zero Maxwell field strength  
with $1/r^2$ asymptotic behavior (like a magnetic monopole).}
These gravitational instantons are known as Abelian Eguchi-Hanson (AEH) instantons\,\cite{Eguchi:1976db}. 
The AEH instantons possess two properties of great importance for our study. First, their action is no longer vanishing but non-zero, and it is controlled not by the Planck scale but by the electromagnetic coupling. Second, the AEH instantons can have a non-zero spin-$\frac{1}{2}$ index.

For these reasons, AEH instantons have been the subject of several studies in relation to the quality problem of the QCD axion.
Before starting our analysis, it is therefore useful to summarize what has already been discussed in the literature. 
First, ref.\,\cite{Deser:1980kc} studied the gravitational analogues of CP violating effects in QCD, 
recognizing that they arise through the addition of terms 
$\theta_{\textrm{em}}F_{\mu\nu}\tilde{F}^{\mu\nu}$ and 
$\theta_{\textrm{grav}}R_{\mu\nu\rho\sigma}\tilde{R}^{\mu\nu\rho\sigma}$
to the Lagrangian density, where $R_{\mu\nu\alpha\beta}$ and $F_{\mu\nu}$ are the curvature tensor and Maxwell field tensor, respectively, with $\tilde{R}^{\mu\nu\rho\sigma}$ and $\tilde{F}^{\mu\nu}$ their Hodge dual.
It was argued that both terms can be non-trivial in spacetimes with sufficiently complex topology.
Refs.\,\cite{Holman:1992ah,Rey:1992wv} demonstrated that AEH gravitational instantons provide a concrete and calculable new source of intrinsic PQ symmetry breaking by quantum gravity. The results of refs.\,\cite{Holman:1992ah,Rey:1992wv} suggest that the gravitational contribution to the axion potential could saturate the current experimental bound on the neutron electric dipole moment.
Ref.\,\cite{Chen:2021jcb} (see also refs.\,\cite{Chen:2021hfq,Chen:2021wcf}) 
considers colored Eguchi-Hanson (CEH) gravitational instantons.
CEH gravitational instantons are the non-abelian analogue of AEH instantons\,\cite{Boutaleb-Joutei:1979vmg,Chakrabarti:1987kz}.   
Ref.\,\cite{Chen:2021jcb} makes the bold claim that CEH gravitational instantons compromise the axion solution to the strong CP problem.

In this work, we will critically reanalyze the role of AEH and CEH gravitational instantons as responsible for the explicit gravitational breaking of the PQ symmetry.
 We anticipate that the effects we will find will be significantly smaller than those previously reported, and we will argue that there is no reason for the PQ symmetry to suffer from a quality problem due to gravity.

\section{Gravitational instantons and the   quality problem of the QCD axion}
\label{sec:GraIn}

We investigate the gravitational contribution to the axion potential $V(a)$ arising from gravitational non-perturbative effects, specifically as the sole sources of CP violation induced by gravity. 

In section\,\ref{sec:PerturbativeGravity}, we will begin by making some perturbative comments regarding the axion's anomalous coupling to gravity, while in section\,\ref{sec:NonPerturbativeGravity} we will discuss the non-perturbative effects.

\subsection{Perturbative considerations}\label{sec:PerturbativeGravity}

The axion, a Goldstone boson (GB) of the non-linearly realized chiral 
$U(1)_{\rm{PQ}}$ symmetry, exhibits a mixed anomaly with QCD and gravity. These anomalies couple the axion to the QCD field strength and its dual, as well as to gravitons through the Riemann tensor and its dual\,\cite{Alvarez-Gaume:1983ihn}. 
The anomalous part of the axion effective Lagrangian reads 
\begin{align}
    \mathcal{L}_a\supset &
    -\frac{1
}{32\pi^2 f_a}\,a\,G^a_{\mu\nu}\tilde{G}^{a,\mu\nu}  
\nn\\
    &-\frac{
E
}{32\pi^2 Nf_a}\,a\,F_{\mu\nu}\tilde{F}^{\mu\nu} \nn\\
    &+ 
\frac{
G
}{
768 \pi^2 N f_a 
}\,a\,R_{\mu\nu\rho\sigma}\tilde{R}^{\mu\nu\rho\sigma}
\,.\label{eq:EffLagrFull2}
\end{align}
where $\alpha_S \equiv g_S^2/4\pi$ (with $g_S$ being the strong coupling), 
$\alpha \equiv e^2/4\pi$ (with $e$ being the fundamental electric charge), while
$N$, $E$ and $G$ are model-dependent anomaly coefficients of order one, which can be computed explicitly for a given UV-complete axion model.  
To provide an example, we calculate the value of these coefficients in the case of a UV-completion of the KSVZ type (cf. 
ref.\,\cite{DiLuzio:2020wdo} for a review).
We consider the generalized KSVZ model described by the  Lagrangian density 
\begin{align}\label{eq:KSVZ}
\mathcal{L}_{\textrm{KSVZ}} = &
(\partial_{\mu}\Phi)^*(\partial^{\mu}\Phi) - 
V(\Phi)\nn\\
&
+ 
\sum_{\mathcal{Q}}\bigg[
\overline{\mathcal{Q}}
i\slashed{D}\mathcal{Q}
- \big(
\lambda_{\mathcal{Q}}\Phi \overline{\mathcal{Q}_L}\mathcal{Q}_R + \textrm{h.c.}\big)\bigg]~,\nn\\
V(\Phi) = & \lambda_{\Phi}
\bigg(
|\Phi|^2 - \frac{v_a^2}{2}
\bigg)^2\,,
\end{align}
with spontaneously broken global $U(1)_{\textrm{PQ}}$ symmetry with order parameter $v_a$.
 We parametrize the complex scalar field as 
$\Phi =
(v_a + \rho)e^{ia/v_a}/\sqrt{2}$,
with the Goldstone field $a$ that plays the role of the axion.  
In full generality, we consider vector-like fermions $\mathcal{Q}$ transforming according to the SM 
representation
\begin{align}
\mathcal{Q} \sim \left(
\mathcal{C}_{\mathcal{Q}},
\mathcal{I}_{\mathcal{Q}},
\mathcal{Y}_{\mathcal{Q}}
\right)_{SU(3)_{\textrm{C}}\times SU(2)_{\textrm{L}} \times U(1)_{\textrm{Y}}}\,.\label{eq:FunTra}
\end{align}
We find 
\begin{align}
N&\equiv \sum_{\mathcal{Q}}
\chi_{\mathcal{Q}}
d(\mathcal{I}_{\mathcal{Q}})
T(\mathcal{C}_{\mathcal{Q}})\,,\nn\\
E &\equiv \sum_{\mathcal{Q}}
\chi_{\mathcal{Q}}d(\mathcal{C}_{\mathcal{Q}})
\textrm{Tr}(q_{\mathcal{Q}}^2)\,,\nn\\
G &\equiv \sum_{\mathcal{Q}}
\chi_{\mathcal{Q}}d(\mathcal{I}_{\mathcal{Q}})
d(\mathcal{C}_{\mathcal{Q}})\,,
\label{eq:FundaCoeff}
\end{align}
with $f_a \equiv v_a/2N$.
$\chi_{\mathcal{Q}} = \pm 1$ with $\chi_{\mathcal{Q}} = +1$ for a Yukawa coupling of the form $\lambda_{\mathcal{Q}}\Phi \overline{\mathcal{Q}_L}\mathcal{Q}_R + \textrm{h.c.}$ and $\chi_{\mathcal{Q}} = -1$ for a Yukawa coupling of the form $\lambda_{\mathcal{Q}}\Phi \overline{\mathcal{Q}_R}\mathcal{Q}_L + \textrm{h.c.}$;
$T(\mathcal{C}_{\mathcal{Q}})$ is the colour Dynkin
index for the representation $\mathcal{C}_{\mathcal{Q}}$ with $SU(3)_{\textrm{C}}$ generators $T^a_{\mathcal{C}_{\mathcal{Q}}}$, and it is defined by the condition $\textrm{Tr}(
T^a_{\mathcal{C}_{\mathcal{Q}}}
T^b_{\mathcal{C}_{\mathcal{Q}}}
) =  T(\mathcal{C}_{\mathcal{Q}})\delta_{ab}$; 
$d(\mathcal{I}_{\mathcal{Q}})$ denotes the dimension of the weak isospin representation $\mathcal{I}_{\mathcal{Q}}$. 
The electric charge operator is defined by  
$q_{\mathcal{Q}} \equiv T^3_{\mathcal{I}_{\mathcal{Q}}} + 
\mathcal{Y}_{\mathcal{Q}}$; notice that 
$\mathcal{Y}_{\mathcal{Q}}$ is to be understood as 
$\mathcal{Y}_{\mathcal{Q}}\, \mathbbm{1}_{d(\mathcal{I}_{\mathcal{Q}}) \times d(\mathcal{I}_{\mathcal{Q}})}$ with $\mathcal{Y}_{\mathcal{Q}}$ a
scalar number. 
It is possible to rewrite $\textrm{Tr}(q_{\mathcal{Q}}^2)  =
\textrm{Tr}
[(T^3_{\mathcal{I}_{\mathcal{Q}}} + 
\mathcal{Y}_{\mathcal{Q}})
(T^3_{\mathcal{I}_{\mathcal{Q}}} + 
\mathcal{Y}_{\mathcal{Q}})] 
= T(\mathcal{I}_{\mathcal{Q}}) 
+ \mathcal{Y}_{\mathcal{Q}}^2 
d(\mathcal{I}_{\mathcal{Q}})$, where $T(\mathcal{I}_{\mathcal{Q}})$ is the Dynkin index of the $SU(2)_{\textrm{L}}$  representation $\mathcal{I}_{\mathcal{Q}}$ and where we used $\textrm{Tr}(T^3_{\mathcal{I}_{\mathcal{Q}}}) = 0$. 
The electromagnetic contribution, which is blind to $SU(3)_{\textrm{C}}$, is proportional 
to the dimension of the  
representation $\mathcal{C}_{\mathcal{Q}}$ and to the trace of the charge 
operator $q_{\mathcal{Q}}^2$. 
On the other hand, the gravitational coupling, which is blind to both $SU(3)_{\textrm{C}}$ and $SU(2)_{\textrm{L}}$, is proportional 
to the dimension of the  
representation $\mathcal{C}_{\mathcal{Q}}$ and $\mathcal{I}_{\mathcal{Q}}$. 
Therefore, the gravitational coupling of the axion does not vanish even in the case in which one takes the vector-like fermions $\mathcal{Q}$ to be completely uncharged under the $SU(3)_{\textrm{C}}\times SU(2)_{\textrm{L}}$ gauge  group (that is, $d(\mathcal{I}_{\mathcal{Q}}) = 
d(\mathcal{C}_{\mathcal{Q}}) = 1$).\\In the infrared---that is, at the level of the chiral axion Lagrangian---a chiral rotation of the light quarks has the following effects: {\it i)} eliminates the effective axion-gluon coupling, {\it ii)} introduces a derivative axial-vector axion-quark coupling, and {\it iii)} modifies the effective axion-photon and axion-graviton couplings. As far as the latter are concerned, we find
\begin{align}
    \mathcal{L}_a\supset 
   & {-\frac{1}{32\pi^2 f_a}}\bigg[
\frac{E}{N} - 
6\textrm{Tr}(Q\mathrm{q}^2)
\bigg]
a F_{\mu\nu}\tilde{F}^{\mu\nu} \nn\\
&+ 
\frac{1}{
768 \pi^2 f_a 
}
\bigg(
\frac{G}{N} - 6
\bigg)\,a\,R_{\mu\nu\rho\sigma}\tilde{R}^{\mu\nu\rho\sigma}\,,\label{eq:NoGlu1}
\end{align}    
with  $\mathrm{q} \equiv \textrm{diag}(2/3,-1/3,-1/3)$ and 
$Q =  
(
m_u^{-1} +
m_d^{-1}+
m_s^{-1})^{-1}
\textrm{diag}(1/m_u,1/m_d,1/m_s)$. 
We note  that 
in the original KSVZ model with one vector-like pair of heavy fermions with 
    $\mathcal{Q} \sim (3,1,0)$ we have $N=1/2$ and  
    $G=3$.
   In this case, therefore, the effective axion-graviton coupling vanishes. 
More in general, in the case in which we have one single heavy fermion, 
we find, whatever $SU(2)_{\textrm{L}}\times U(1)_{\textrm{Y}}$ quantum numbers, $G/N = d(\mathcal{C}_{\mathcal{Q}})/T(\mathcal{C}_{\mathcal{Q}})$. 
This immediately implies that in the case of  the fundamental representation of $SU(3)_{\textrm{C}}$ for which  
$d(\mathcal{C}_{\mathcal{Q}}) = 3$ and $T(\mathcal{C}_{\mathcal{Q}}) = 1/2$ we have $G/N = 6$ and the gravitational coupling vanishes. 
However, in more general 
   KSVZ models we have 
   $G/N \neq 6$.  
For instance, in the non-minimal solution with 
    $\mathcal{Q}_1\sim (8,1,0)$,
    $\mathcal{Q}_2\sim (6,1,0)$
and opposite $\chi_{\mathcal{Q}}$ charge (that is, the non-minimal model that gives $N_{\textrm{DW}}  = 1$) we get $G/N = 4$. 
Furthermore, following from the previous argument, in the case of a single $\mathcal{Q}$uark, $G/N \neq 6$
if $\mathcal{Q}$ transforms according to 
$SU(3)_{\textrm{C}}$
irreps of dimension higher than the fundamental.  To fix ideas, we consider some of the 
most relevant cases discussed in ref.\,\cite{DiLuzio:2016sbl}. 
\begin{table*}[!htb!]
\renewcommand{\arraystretch}{1.4}
\begin{center}
\begin{adjustbox}{max width=1\textwidth}
\begin{tabular}{||c||c|c|c|c|c|c|c|c|c|c||}
\hline\hline
 \multirow{2}{*}{$\mathcal{Q}$ representation} & $R_1$ & 
 $R_3$ & $R_6$ &  
 $R_8$ & 
 $R_9$ & $R_{12}$  & \multirow{2}{*}{$R_8 \ominus R_3$} & \multirow{2}{*}{$R_8 \oplus R_6 \ominus R_9$} 
  & \multirow{2}{*}{$R_6 \oplus  R_9$}   
  & \multirow{2}{*}{$R_{12} \ominus R_9$} \\ 
 & $(3, 1,-1/3 )$ & $(3, 2,1/6 )$ & $(3, 3,-1/3 )$ & $(3, 3,-4/3 )$ & $(\bar{6},1,-1/3 )$ & $(8,1,-1)$ & & & & \\ \hline
$E/N$  &2/3  & 5/3 & 14/3 & $44/3$ & $4/15$ & $8/3$ & $122/3$ & $170/3$ & $23/12$ & $44/3$  \\ \hline
$G/N$  & 6 & 6 & 6 & 6 & 12/5 & 8/3 & 6 & 24 & 15/4 & 4 \\ \hline
$N_{\textrm{DW}}$ 
& 1 & 2 & 3 & 3 & 5 & 6 & 2 & 2 & 4 & 1  \\ 
 \hline\hline
\end{tabular}
\end{adjustbox}
\end{center}
\vspace{-0.5cm}
\caption{{\it 
Numerical values for 
$E/N$, $G/N$ and $N_{\textrm{DW}} = 2N$  for some of the most relevant 
$\mathcal{Q}$ representations  discussed in ref.\,\cite{DiLuzio:2016sbl}.\footnote{The axion is defined as an angular variable over the domain
$[0,2\pi v_a)$.
On  the contrary, 
the QCD-induced axion potential is periodic in $[0,2\pi f_a)$ with $f_a \equiv v_a/2N$ defined before.  
It is customary  to define 
 the domain wall (DW) number $N_{\textrm{DW}} \equiv 2N$; 
 this is  the number of inequivalent degenerate minima of the axion potential (generated by QCD). 
 $N_{\textrm{DW}}$ is crucially related to the  so-called domain wall problem\,\cite{Sikivie:1982qv}.
 We distinguish between two cases.
    {\it i)}
    $N_{\textrm{DW}} = 1$ (no remnant discrete symmetry). 
    There is no domain wall problem in this case. 
    {\it ii)} $N_{\textrm{DW}} \geqslant 2$.
    In this case, domain walls are stable.
 }
The symbol $\oplus$ refers to $\chi_{\mathcal{Q}} = +1$ while $\ominus$ to $\chi_{\mathcal{Q}} = -1$.
}}\label{tab:TableRep}
\end{table*}
In table\,\ref{tab:TableRep}, we show for which variations of the KSVZ model it is possible to obtain a non-zero value for the gravitational coupling $G/N - 6$. 
When non-zero, we observe that the value of the combination $G/N - 6$ remains of order one in realistic situations.

From a phenomenological standpoint, it is clear that the gravitational coupling of the QCD axion is of negligible relevance in perturbation theory. 
 In the weak–field limit, the quantization of gravity amounts to expanding
the metric around the Minkowski background
$g_{\mu\nu} = \eta_{\mu\nu}  + 
\kappa h_{\mu\nu}$.
The inclusion of the factor $\kappa \equiv \sqrt{32\pi G_N}$ in the definition of
the graviton field $h_{\mu\nu}$ gives this field a mass dimension of unity and results in a kinetic term (arising from the standard Einstein-Hilbert action) with standard normalization.
For matter interactions, the order of $\kappa$ keeps
track of the number of gravitons involved in an interaction. 
Consequently, for a Lagrangian 
density of the form
$\mathcal{L} = y\,a\,R_{\mu\nu\rho\sigma}\tilde{R}^{\mu\nu\rho\sigma}$ we find 
\begin{align}
&\hspace{2cm}\begin{adjustbox}{max width=0.94\textwidth}
\raisebox{-14.5mm}{
	\begin{tikzpicture}
\draw[thick,dashed][thick] (-1,0)--(0,0);
\draw[black,style={decorate, decoration={snake,segment length=3mm,amplitude=.6mm}},double][thick] (0,0)--(1.25,1); 
\draw[black,style={decorate, decoration={snake,segment length=3mm,amplitude=.6mm}},double][thick] (0,0)--(1.25,-1); 
\draw[black,fill=tangerineyellow,thick] (0,0)circle(3pt); 
\draw[black,->,>=Latex] (0.8,-0.45)--(1.2,-0.7);
\draw[black,->,>=Latex] 
(0.8,+0.45)--(1.2,+0.7);
\node at (1.2,0.42) {\scalebox{1}{$p_1$}}; 
\node at (1.2,-0.42) {\scalebox{1}{$p_2$}}; 
\node at (0.75,1.25) {\scalebox{1}{{\color{firebrick}{$\mu\nu,\lambda$}}}}; 
\node at (0.75,-1.25) {\scalebox{1}{{\color{firebrick}{$\rho\sigma,\lambda^{\prime}$}}}};
	\end{tikzpicture}}
 \label{eq:Loop2Graveff}   
  \end{adjustbox} \hspace{-0.275cm}
\nn\\&= 2yi\kappa^2\varepsilon_{
\mu\rho\alpha\beta 
 }p_1^{\alpha}p_2^{\beta}
\big[
p_{1\,\sigma}p_{2\,\nu} - 
(p_1\cdot p_2)\eta_{\nu\sigma}
\big]\,,
\end{align}
while the interaction Lagrangian density $\mathcal{L} 
= k\,a\,F_{\mu\nu}\tilde{F}^{\mu\nu}$ generates the amplitude 
\begin{align}
\begin{adjustbox}{max width=0.94\textwidth}
\raisebox{-13.5mm}{
	\begin{tikzpicture}
\draw[thick,dashed][thick] (-1,0)--(0,0);
\draw[black,style={decorate, decoration={snake,segment length=3mm,amplitude=.6mm}}][thick] (0,0)--(1.25,1);  
\draw[black,style={decorate, decoration={snake,segment length=3mm,amplitude=.6mm}}][thick] (0,0)--(1.25,-1); 
\draw[black,fill=brilliantlavender,thick] (0,0)circle(3pt); 
\draw[black,->,>=Latex] (0.8,-0.4)--(1.2,-0.7);
\draw[black,->,>=Latex] 
(0.8,+0.4)--(1.2,+0.7);
\node at (1.2,0.42) {\scalebox{1}{$p_1$}}; 
\node at (1.2,-0.42) {\scalebox{1}{$p_2$}}; 
\node at (0.75,1.25) {\scalebox{1}{{\color{firebrick}{$\mu,\lambda$}}}}; 
\node at (0.75,-1.25) {\scalebox{1}{{\color{firebrick}{$\nu,\lambda^{\prime}$}}}};
	\end{tikzpicture}}~ 
  \end{adjustbox}
\equiv 4ik\varepsilon_{\mu\nu\rho\sigma}p_1^{\rho}p_2^{\sigma}\,.
\end{align}
The axion-graviton coupling, if compared to the axion-photon one, 
pays a huge suppression factor of the order of 
\begin{align}
\frac{a\to\textrm{gravitons}}{
a\to \textrm{photons}
} \sim \alpha^{-1}\left(\frac{m_a}{M_{\textrm{Pl}}}\right)^2\,,\label{eq:GapMas}
\end{align}
where we have treated powers of momentum  as the axion rest mass $m_a$.
Based on dimensional analysis, 
the decay width of the axion into two gravitons can be estimated to be of the order of
\begin{align}
\Gamma_{a\to gg} \approx \frac{G_N^2 m_a^7}{\pi^4 f_a^2}\,,\label{eq:StimaVitaMedia}
\end{align}
which is phenomenologically irrelevant.

From a model-building perspective, one might consider attempting to engineer a mechanism that leads to an amplification of the gravitational coupling of the axion (without altering that of QCD). 
A potentially interesting possibility is to implement a clockwork mechanism\,\cite{Choi:2015fiu,Kaplan:2015fuy}, along the lines of that proposed in ref.\,\cite{Farina:2016tgd}. 
In its original formulation, the clockwork mechanism is described by a renormalizable theory involving a sequence of $\mathcal{N}+1$ complex scalar fields $\phi_i$ with a global $U(1)^{\mathcal{N}+1}$ 
symmetry that is spontaneously broken at the scale $f$. 
This global symmetry is further explicitly broken, but in a manner that leaves a residual $U(1)$ symmetry intact. The corresponding NG boson, which can be identified with the QCD axion as discussed in ref.\,\cite{Farina:2016tgd}, resides in a compact field space, where its effective decay constant is determined by $ 3^{\mathcal{N}}f \gg f$. 
The core idea presented in ref.\,\cite{Farina:2016tgd} is as follows: new vector-like fermions responsible for generating the color anomaly are coupled to the final site $\mathcal{N}$ of the scalar chain. This setup still resolves the strong CP problem but with the key distinction that the scale $3^{\mathcal{N}}f$ can be significantly larger than the fundamental symmetry-breaking scale $f$. 
In the standard QCD axion model, the vector-like fermions that mediate the QCD anomaly also contribute to the axion-photon coupling. However, in the clockwork QCD axion framework, additional electromagnetically charged vector-like fermions are coupled to a site $\mathcal{M}<\mathcal{N}$ in the scalar chain. These fermions generate the axion-photon coupling, which is effectively decoupled from the solution to the strong CP problem.
In its simplest realization, the clockwork axion model requires the existence of a single
vector-like colored fermion in the fundamental representation of $SU(3)_{\textrm{C}}$ and a single color neutral vector-like fermion with unit hypercharge and singlet under 
$SU(2)_{\textrm{L}}$.

In our case, it is sufficient to introduce a single heavy vector-like fermion, even if  completely neutral under the Standard Model gauge group, to contribute to the gravitational axion coupling.
The setup we have in mind, therefore, takes the structure  given by the following schematic
\begin{align}
	\begin{tikzpicture}
	 {\scalebox{1}{
    \draw[thick,dashed] (-3.5,0)--(3.5,0);
    \draw[thick] (-3.5,0)--(-2,0);
    \draw[thick] (2,0)--(3.5,0);
    \draw[thick] (-1,0)--(1,0);
    \draw[thick] (3.5,0)--(3.5,0.4);
    \draw[thick] (0,0)--(0,0.4);
    \draw[black,fill=tangerineyellow,thick] (-3.5,0)circle(2.5pt);
    \draw[black,fill=tangerineyellow,thick] (-2.5,0)circle(2.5pt);    
    \draw[black,fill=tangerineyellow,thick] (2.5,0)circle(2.5pt);    
    \draw[black,fill=tangerineyellow,thick] (3.5,0)circle(2.5pt);  
    \draw[black,fill=tangerineyellow,thick] (0,0)circle(2.5pt);      
    \node at (-3.5,-0.5) {\scalebox{1}{$\phi_0$}};
    \node at (-2.5,-0.5) {\scalebox{1}{$\phi_1$}}; 
    \node at (2.5,-0.5) {\scalebox{1}{$\phi_{\mathcal{N}-1}$}};    
    \node at (+3.5,-0.5) {\scalebox{1}{$\phi_{\mathcal{N}}$}}; 
    \node at (0,-0.5) {\scalebox{1}{$\phi_{\mathcal{M}}$}};  
    \node at (+3.5,+0.65) {\scalebox{1}{$\mathcal{Q} \sim \left(
\mathcal{C}_{\mathcal{Q}},
\mathcal{I}_{\mathcal{Q}},
\mathcal{Y}_{\mathcal{Q}}
\right)$}};
    \node at (0,+0.65) {\scalebox{1}{$\Psi \sim \left(
1,
1,
0
\right)$}};
    }}
	\end{tikzpicture}\nn
\end{align}
in which one vector-like fermions $\mathcal{Q}$ is coupled to the last site of the clockwork chain while a second vector-like fermion $\Psi$, uncharged under the Standard Model gauge group, is coupled to the site 
$\mathcal{M}$.  
The effective coupling of the axion to gluons and gravitons arises by performing a chiral rotation to remove the axion from the Yukawa coupling involving the fermions $\mathcal{Q}$ and $\Psi$.  
We find
\begin{align}
\mathcal{L}_a&\supset
    -\frac{1}{32\pi^2f_a}\,a\,G^a_{\mu\nu}\tilde{G}^{a,\mu\nu}\nn\\&  + 
\frac{
G
}{
768 \pi^2 N f_a 
}\,a\,R_{\mu\nu\rho\sigma}\tilde{R}^{\mu\nu\rho\sigma}\,,
\end{align}
with
\begin{align}
\frac{G}{N} =  
\frac{
d(\mathcal{I}_{\mathcal{Q}})
d(\mathcal{C}_{\mathcal{Q}}) 
+
3^{\mathcal{N}-\mathcal{M}}
d(\mathcal{I}_{\Psi})
d(\mathcal{C}_{\Psi}) 
}{
d(\mathcal{I}_{\mathcal{Q}})
T(\mathcal{C}_{\mathcal{Q}})
}\,.
\end{align}
Even considering a fermion $\mathcal{Q} \sim (3,1,0)$ in the color fundamental and isospin singlet, we find that, at low energies, the gravitational coupling of the axion is non-zero and is given by
\begin{align}
  \frac{G}{N} - 6 = 
2\times 3^{\mathcal{N}-\mathcal{M}}\,,
\end{align}
where we used $d(\mathcal{I}_{\Psi}) = 
d(\mathcal{C}_{\Psi}) = 1$. 
The relation between  the fundamental symmetry breaking scale $f$ and the effective decay constant $f_a$ is given by $f_a = 3^{\mathcal{N}}f$. 
Furthermore, we note that in this particular realization with $\mathcal{Q} \sim (3,1,0)$ and $\Psi \sim (1,1,0)$, the axion-photon coupling is entirely generated at low energies and is therefore controlled by the effective axion decay constant $f_a$.

We have thus found that the coupling to gravitons can be exponentially large, and its magnitude is controlled by the distance between the site coupled to the colored fermions and the site coupled to the uncharged ones.  
It is natural at this point to ask how large this enhancement can be. 
It is not difficult to realize that, at least within the context of this minimal setup, there is a fundamental obstruction to the attempt to make the gravitational coupling arbitrarily large. 
On the one hand, we would like to consider an axion mass as large as possible in order to minimize the ratio given by eq.\,(\ref{eq:GapMas}). 
On the other hand, the axion mass generated by non-perturbative QCD effects is described by the scaling\,\cite{GrillidiCortona:2015jxo}
\begin{align}
m_a \approx 
5.7\times 10^{-10}
\left(
\frac{10^{16}\,\textrm{GeV}}{f_a}
\right)\textrm{eV}\,.
\end{align}
Consequently, increasing the axion mass implies decreasing the value of $f_a = 3^{\mathcal{N}}f$, and thus considering a small value of $3^{\mathcal{N}}$ (assuming a fixed value of $f$, for instance $f=O(100)$ GeV, which approximately corresponds to the smallest possible value, given that at the scale $f$ the massive scalar gears produced by the clockwork mechanism are found). 
However, this latter factor is precisely what should contribute to the enhancement of the gravitational coupling.

To be a bit more concrete, if we consider a GUT-scale value $f_a \approx 10^{16}$ GeV and a spontaneous breaking scale $f\approx 100f$ GeV, we obtain an enhancement factor (we take $\mathcal{M} = 0$ to maximize the effect) $3^{\mathcal{N}} \approx 10^{14}$, which, however, is still too small to make the gravitational coupling phenomenologically relevant. 
Numerically, eq.\,(\ref{eq:StimaVitaMedia}) gives 
\begin{align}
\Gamma_{a\to gg}^{-1} &\approx 
10^{102}\left(
\frac{f_a}{10^{16}\,\textrm{GeV}}
\right)^9 3^{-2\mathcal{N}}\,\textrm{s} 
\nn\\&\gg \tau_{\textrm{U}} \approx 
10^{17}\,\textrm{s}\,,
\end{align}
where $\tau_{\textrm{U}}$ is the age of the universe.\\In perturbation theory, since $G\tilde{G} \equiv G^a_{\mu\nu}\tilde{G}^{a,\mu\nu}$ is a total derivative, the axion Lagrangian is shift-invariant. However, because of QCD non-perturbative effects, $G\tilde{G}$ will generate an axion potential. Analogously, $R\tilde{R} \equiv R_{\mu\nu\rho\sigma}\tilde{R}^{\mu\nu\rho\sigma}$, is a total derivative and, \textit{in perturbation theory}, it does not spoil the axion shift-invariance. 
We now move to consider non-perturbative effects.

\subsection{Non-perturbative breaking induced by gravity}\label{sec:NonPerturbativeGravity}

\begin{figure}[h!]
\begin{center}
    \includegraphics[width=.5\textwidth]{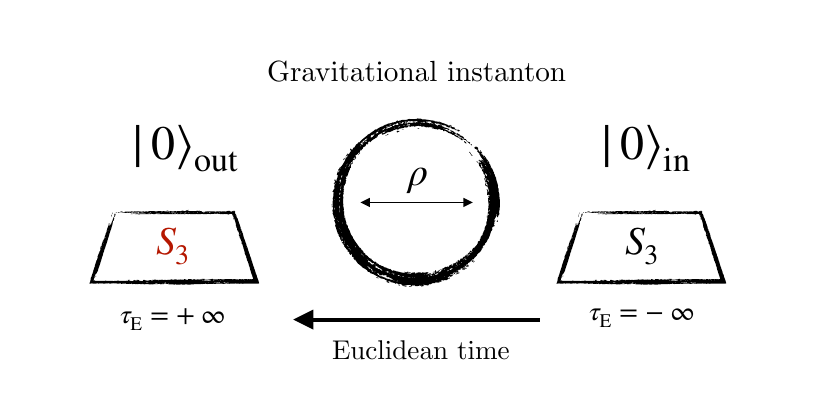}
    \caption{\em Schematic representation of a gravitational instanton of size $\rho$, as a tunneling event between two disjoint vacua, intended as two flat Euclidean spacetimes.}
    \label{Fig:Gravitational Instanton}
\end{center}
\end{figure}

We ask whether gravitational non-perturbative effects can induce $R\tilde{R}$ to generate an axion potential. This could be naturally interpreted as the axion potential induced by gravitational interactions. Mimicking the strong interactions\,\cite{JackiwRebbi,CallanDashenGross,Jackiw, huang1992quarks}, where in the Euclidean functional integral $\mathcal{Z}_{\rm{E}}$ we have to sum over all the topologically distinct gauge sectors, we have the full (Euclidean) generating functional\,\cite{HAWKING1978349}
\begin{align}
\mathcal{Z}_{\rm{E}}&\equiv\lim_{\tau_{\rm{E}}\rightarrow\infty}\tensor[_{\rm{out}}]{\bra{0}e^{-\hat{H}\tau_{\rm{E}}}\ket{0}}{_{\rm{in}}}\nn\\
&=\int\left[\mathcal{D}\mathcal{A}\right]\left[\mathcal{D}\Psi\right]\left[\mathcal{D}g\right]e^{-S_{\rm{E}}}\,,\label{eq:VPA}
\end{align}
where, $[\mathcal{D}\Psi]$ means summing over all the matter fields in the theory, $[\mathcal{D}\mathcal{A}]$ over all the topologically distinct sectors of the Yang-Mills theory and $[\mathcal{D}g]$ over all the gravitational sectors. 
At this stage, it is crucial to pause for an important remark. Regarding the gravitational part of the functional integral defining the vacuum persistence amplitude in eq.\,(\ref{eq:VPA}), one should sum over all 4-dimensional Riemannian spaces $\{M,g_{\mu\nu}\}$
with arbitrary topology for the manifold 
$M$ and arbitrary metric $g_{\mu\nu}$, subject to the condition that $M$ has a prescribed boundary $\partial M$, and that $g_{\mu\nu}$ induces some prescribed metric $k_{ab}$ on $\partial M$.
It seems plausible that the appropriate boundary condition for the vacuum persistence amplitude corresponds to metrics that are asymptotically Euclidean, with a boundary at infinity that can be interpreted as a 3-sphere equipped with its standard metric. 
We could thus attempt to imagine the contribution of a gravitational instanton as schematized in 
fig.\,\ref{Fig:Gravitational Instanton}.
 
From an operational standpoint, this sum can be organized as follows\,\cite{Gibbons:1978ac}. First, it is possible to divide the positive definite metrics $g_{\mu\nu}$ on $M$, which induce the given metric $k_{ab}$ on the boundary $\partial M$, into equivalence classes under conformal transformations with a conformal factor $\Omega$ that is positive and equal to one on $\partial M$. Second, it is possible to select a metric from each conformal equivalence class that satisfies $R = 0$. One can then explicitly perform the integration over conformal deformations of this metric. 
 This leaves the task of summing over all metrics that satisfy $R=0$. 
 In full generality, 
 the action of these metrics is given by a boundary term (the bulk term vanishes).
 
 At this point, in order for the functional integration to behave properly, one can invoke the so-called positive-action conjecture\,\cite{Gibbons:1978ac,Page:1978zz}: 
all non-singular asymptotically flat metrics with $R = 0$ everywhere have non-negative action.  
It is further conjectured that the action is zero if and only if the space is flat. 
The validity of the positive action conjecture---which, we reiterate, must hold true in order to give meaning to the functional integration---has a striking consequence of great relevance to our discussion. In fact, the positive action conjecture requires that there are no Ricci-flat asymptotically Euclidean metrics other than flat space\,\cite{Page:1978zz,Gibbons:1979xn}. 
Consequently, if we restrict ourselves to considering exclusively metrics that are asymptotically Euclidean, it would not be legitimate to include gravitational instantons  (as defined in section\,\ref{sec:GI}) in the sum defining the vacuum persistence amplitude. 
This argument would resolve our issue at its root. The gravitational instantons discussed in refs.\,\cite{Holman:1992ah,Rey:1992wv,Chen:2021jcb}, which could potentially spoil the axion solution to the strong CP problem, simply must not be considered in the gravitational path integral.

However, in the remainder of our analysis, we choose to follow the discussion in ref.\,\cite{Gibbons:1979xn}, in which an alternative asymptotic condition was proposed by introducing a fundamental group at infinity: the metric locally approaches the Euclidean one but a neighborhood of infinity has the topology of 
$(S^3/\Gamma)\times \mathbb{R}$ where $\Gamma$ is a finite group of isometries acting on $S^3$. 
These are nothing but the ALE spaces discussed in section\,\ref{sec:GI}.

In the following, therefore, we will contemplate summing over ALE spaces. 
The reasoning behind this choice is to demonstrate, as we will explicitly do, that even in the case where ALE spaces are taken into consideration, the low-energy physics that is potentially sensitive to these geometries remains protected, even in the absence of Planck-scale suppression. 
In particular, as explained in section\,\ref{sec:GI}, we will focus on the contribution of
 the AEH gravitational instantons.
 
We have, in full generality, the Euclidean action 
\begin{align}
S_{\rm{E}}=S_{\rm{grav}}+S_{\rm{gauge}}+S_{\rm{matter}}+S_{\theta}\,,\label{eq:FullAction}
\end{align}
in which
\begin{equation}
\begin{split}
    S_{\rm{grav}}=&
    -\frac{1}{16\pi G_N}
    \int_{{M}} d^4x\,\sqrt{g}\,R\\
    &
    -\frac{1}{8\pi G_N}\int_{\partial M}
    d^3\vec{x}
    \,\sqrt{k}\,[K]
    \,,\label{eq:EHPlusBouns}
\end{split}
\end{equation}
where the volume integral of the Ricci scalar $R$ is taken over $M$ with boundary $\partial M$
and $[K]\equiv K-K_0$ denotes the difference between the trace of the second fundamental form of the boundary $\partial M$ in the metric $g_{\mu\nu}$ (denoted by $K$), and its value in the flat metric (denoted by $K_0$); 
$k$ is the determinant of the induced metric on the boundary.\\
The second term in eq.\,(\ref{eq:EHPlusBouns}), known as the Gibbons-Hawking-York boundary term, is an important term added to the Einstein-Hilbert action in the context of General Relativity (GR) when spacetime has boundaries, and is necessary because the variation of the Ricci scalar $R$, which contains second derivatives of the metric, leads to boundary terms involving first derivatives of the metric that must be canceled to ensure a well-posed variational principle\,\cite{York:1971hw,York:1972sj,Gibbons:1976ue}. 
In eq.\,(\ref{eq:FullAction}), we further have
\begin{align*}
    S_{\rm{gauge}}+S_{\rm{matter}}&=\int_{{M}} d^4x\,\sqrt{g}\,\bigg\{\frac{1}{4\,g_{S}^2}G_{\mu\nu}^aG^{a,\mu\nu}\\&+\frac{1}{4\,e^2}\,F_{\mu\nu}F^{\mu\nu}+\mathcal{L}[\mathcal{A},\Psi]\bigg\}\,,
\end{align*}
where $\mathcal{L}[\mathcal{A},\Psi]$, in which $\mathcal{A}$ and $\Psi$ collectively representing the gauge and matter fields, respectively, and includes the axion Lagrangian $\mathcal{L}_a$. Finally, we have the topological terms (that are purely imaginary in Euclidean spacetime\,\cite{SHIFMAN198046}) 
\begin{align}\label{eq:Topological terms}
    S_\mathcal{\theta}=i\int_{{M}} d^4x\,\sqrt{g}\,\bigg\lbrace&\frac{\overline{\theta} }{32\pi^2}G_{\mu\nu}^{a}\tilde{G}^{a,\mu\nu}+\frac{\theta_{\rm{em}} }{32\pi^2}F_{\mu\nu}\tilde{F}^{\mu\nu}\nn\\
    &-\frac{\theta_{\rm{grav}}}{768\pi^2}R_{\mu\nu \rho\sigma}\tilde{R}^{\mu\nu \rho\sigma}\bigg\rbrace\,,
\end{align}
where $\theta_{\rm{em}}$ is non-zero in non-topologically trivial backgrounds. In the following we will demonstrate that AEH instantons\,\cite{Eguchi:1976db} (see Appendix \ref{Eguchi-Hanson metric}) allow for an explicit and controlled computation of the gravitationally induced axion potential, identifying it as the dominant contribution from non-perturbative gravitational effects. Indeed, in the case of gravitational instantons, the one-instanton contribution to the Euclidean path integral will be
\begin{align}\label{eq:Zgrav 1 inst}
\left.\mathcal{Z}_{\rm{E}}\right|_{\rm{1-inst}}&=\sum_{\substack{\overline{g}=\overline{g}_{\rm{AEH}}}}\int\left[\mathcal{D}\mathcal{A}\right]\left[\mathcal{D}\Psi\right]\left[\mathcal{D}h\right]e^{-S_{\rm{E}}[\mathcal{A},\Psi,\overline{g},h]}\,,\nn\\
g_{\mu\nu}&=\overline{g}_{\mu\nu}+h_{\mu\nu}/M_{\rm{Pl}}\,.
\end{align}
The AEH instantons have Euclidean action (see appendix\,\ref{Eguchi-Hanson metric})
\begin{equation}
S_{\rm{E}}\left[\overline{g}_{\rm{AEH}}\right]=\frac{\pi  p^2}{\alpha}-i\,\frac{p^2}{2} \theta _{\text{em}}-i\frac{\theta_{\rm{grav}}}{16}\,,
\end{equation}
where $e$ is a $U(1)$ gauge charge that may naturally be interpreted as the electric charge and $p$ is a real number. Consequently, eq.\,\eqref{eq:Zgrav 1 inst} becomes
\begin{align}\label{eq:Zgrav 1 inst, Eguchi Hanson}
\left.\mathcal{Z}_{\rm{E}}\right|_{\rm{1-inst}}&=\sum_{p}\int\left[\mathcal{D}\mathcal{A}\right]\left[\mathcal{D}\Psi\right]\left[\mathcal{D}h\right]\times\nn\\
&\times e^{-\frac{\pi  p^2}{\alpha}+i\,\frac{p^2}{2} \theta _{\text{em}}+i\frac{\theta_{\rm{grav}}}{16}+\dots}\nn\\&+\mathcal{O}(e^{-M_{\rm{Pl}}^2\rho^2})\,,
\end{align}
where the ellipses stand for quantum fluctuations around the classical solution. 
At this point, after discussing the theoretical foundations of the semi-classical gravity approach within which we are operating, we can proceed to the calculation of the axion potential generated by gravitational instantons.

\subsection{Contribution to the axion potential}

In full analogy to the computation of the axion potential arising from QCD instanton effects, we will employ the method of 't Hooft operators to compute the axion potential induced by gravitational effects. In ref.\,\cite{PhysRevLett.37.8}, 't Hooft showed that (see appendix\,\ref{Appendix 't Hooft}), at distances much larger than the instanton size---so that we can describe one instanton event in a local way---BPST instantons\,\cite{BELAVIN197585} produce effective interactions for a quark $q^s_{i}$ with color index $i=1,\dots,N_c$ and flavor index $s=1,\dots,N_f$ with effective Lagrangian 
\begin{align}\label{eq:'t Hooft operator QCD 1}
    \Delta\mathcal{L}_{\rm{inst}}(x)&=c\,e^{i\overline{\theta}}\,\left(\mathcal{K}_{N_c}^{(N_f)}\right)^{i_1i_1\dots i_{Nf}}\times\nn\\
    &\times\det_{s,t}\left[{\bar{q}_R^s(x)q^t_L(x)}\right]_{i_1i_2\dots i_{2N_f}}+\rm{h.c.}\,,
\end{align}
where 
\begin{equation}
c=O(1)\exp[-S_{\rm{E}}(\mathcal{A}_{\textrm{inst}})]\,.
\end{equation} 
We indicate with $\mathcal{A}_{\textrm{inst}}$ the classical BPST instanton\,\cite{BELAVIN197585} whose action is given by $S_{\rm{E}}(\mathcal{A}_{\textrm{inst}})=8\pi^2/g_S^2$ with $g_S$ the strong coupling constant; in eq.\,(\ref{eq:'t Hooft operator QCD 1}), $\mathcal{K}_{N_c}^{(N_f)}$ is a tensor of the color group $SU(N_c)$.

The physics content of eq.\,(\ref{eq:'t Hooft operator QCD 1}) can be intuitively understood from the celebrated index theorem\,\cite{Weinberg:1996kr}, which states that the violation of the chiral charge
\begin{equation*} Q_5=\sum_{j=1}^{N_f}\int d^4x\,\overline{\Psi}_j\gamma_0\gamma_5\Psi_j\,,
\end{equation*} 
in a background $\mathcal{A}$ is $\Delta Q_5=2(n_+-n_-)=4N_f T(\mathcal{C}_{\Psi})n$, where $n$ is the winding number, normalized as
\begin{align*}
n&=\frac{1}{32\pi^2}\int\,d^4x\,\mathcal{G}_{\mu\nu}^a\tilde{\mathcal{G}}^{a,\mu\nu}=0,\pm1,\pm2,\dots\,,
\end{align*}
$\mathcal{G}_{\mu\nu}^a=G_{\mu\nu}^a(\mathcal{A})$
(that is, $\mathcal{G}_{\mu\nu}^a$ denotes $G_{\mu\nu}^a$ evaluated on the gauge configuration $\mathcal{A}$) and $n_+-n_-=\rm{index}\left(i\slashed{\mathcal{D}}\right)_{\mathcal{A}}$ is the index of the Dirac operator in the background of $\mathcal{A}$, \textit{i.e.} the number of normalizable eigenvectors with zero eigenvalue of $i\slashed{\mathcal{D}}_{\mathcal{A}}$ (the so-called \textit{zero modes}). Under a chiral transformation, we have $q_R\rightarrow q_R e^{-i\alpha}$, $q_L\rightarrow q_L e^{i\alpha}$. Furthermore, for simplicity, we have assumed that all the fermions $\Psi_j$'s are in the same representation of the color group with Dynkin index $T(\mathcal{C}_{\Psi})=1/2$. From the index theorem, in the background of one instanton (for which $n=1$), the chiral charge is violated by $\Delta Q_5=2N_f$ units. That is to say, each flavor violates the chiral charge by $2$ units. This effect of non-perturbative chiral charge violation in presence of one instanton event can be incarnated by the effective vertex in eq.\,(\ref{eq:'t Hooft operator QCD 1}), which indeed violates, in a non-perturbative way, the chiral charge by $2N_f$ units.

In the PQ solution\,\cite{PhysRevLett.38.1440}, $\overline{\theta}$ is promoted to $\overline{\theta}\rightarrow\overline{\theta}+2N a/v_a$; therefore, eq.\,\eqref{eq:'t Hooft operator QCD 1} is modified to (with $f_a \equiv v_a/2N$, and omitting the color indices $i$)
\begin{align}\label{eq:'t Hooft operator QCD 2}
    \Delta\mathcal{L}_{\rm{inst}}(x)&=c\,e^{i\overline{\theta}}e^{i a/f_a}\,\mathcal{K}_{N_c}^{(N_f)}\det_{s,t}\left[{\bar{q}_R^s(x)q^t_L(x)}\right]\nn\\
    &+\rm{h.c.}\,.
\end{align}
One then can easily convince oneself that, closing the fermion legs using all the interactions at disposal contained in $S_{\rm{matter}}+S_{\rm{gauge}}$, obtains the instanton contribution to the axion potential; indeed eq.\,(\ref{eq:'t Hooft operator QCD 2}) is not invariant under the shifts $a(x)/f_a\rightarrow a(x)/f_a+\alpha$.

Following the methodology applied in the QCD case, we construct the gravitational analogue of 't Hooft vertex in eq.\,\eqref{eq:'t Hooft operator QCD 1} by analyzing the Dirac zero-modes, \textit{i.e.} the violation of the chiral charge, in the background of a gravitational instanton. That is to say, we have to solve for \textit{normalizable} solutions of the covariant and massless Dirac equation:
\begin{equation}\label{eq:zero mode equation}
\gamma^\mu\mathcal{D}_{\mu}\psi_f=0\,,
\end{equation}
where $\mathcal{D}_\mu$ is the Dirac operator evaluated in the gravitational instanton metric background twisted by the abelian gauge field $\mathcal{A}_\mu$, that is to say
\begin{align*}
        \gamma^\mu \mathcal{D}_{\mu,\mathcal{A}}&=\slashed{\mathcal{D}}_\mathcal{A}\nn\\
        &=\gamma^a E^\mu_a\bigg(\partial_\mu-\frac{1}{8}\tensor{\omega}{_\mu^{bc}}\left[\gamma^b,\gamma^c\right]+iQ^f_{\rm{em}}\mathcal{A}_\mu\bigg)\,.
\end{align*}
 
\begin{figure}[h!]
\begin{center}
    \includegraphics[width=.5\textwidth]{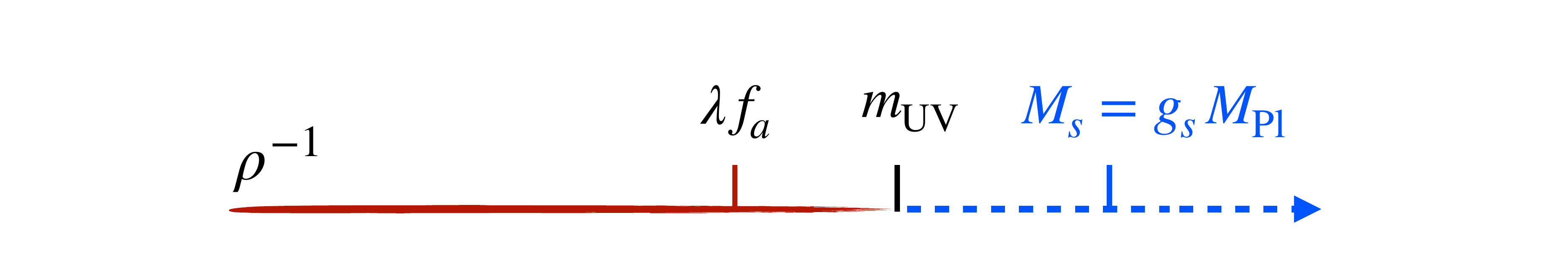}
    \caption{\em The AEH solution is reliable up to some mass scale $m_{\rm{UV}}$, after which corrections to GR become to be relevant due to the onset of particles of the UV-completion.}
    \label{Fig:scales}
\end{center}
\end{figure}

 An explicit computation (see appendix\,\ref{Computation of Dirac zero-modes on Eguchi-Hanson manifold} for details), shows that, differently from what was found in ref.\,\cite{THOOFT1989517} and used in ref.\,\cite{Holman:1992ah}, the number of zero modes in the AEH background is, instead\,\cite{Franchetti_2018}
\begin{align}
    \vert n_+-n_-\vert&=\sum_{2j+1=1}^{[2|Q^f_{\rm{em}}p|+1]}(2j+1)\,,
\end{align}
in which $Q_{\rm{em}}^f$ is the electric charge of the fermion $\psi_f$ in eq.\,\eqref{eq:zero mode equation} and $[\mathcal{N}]$ is the largest integer strictly smaller than $\mathcal{N}$ (for instance $[3]=2$, etc). From topological arguments, see eq.\,\eqref{eq:monopole quantization}, for a fermion in a AEH background, we have the Dirac string quantization condition
\begin{equation}
|Q_{\rm{em}}p|=0,\,1,\,2\,\dots\,.
\end{equation} Therefore, since $Q_{\rm{em}}p$ is an integer, one AEH instanton event violates the chiral charge $Q_5$, associated with the axial current $J^\mu_5=\sum_f\overline{\psi}_f\gamma^\mu\gamma_5\psi_f$, by
\begin{equation}\label{eq:Chiral charge}
\vert \Delta Q_5\vert=2\vert n_+-n_-\vert=2|Q^f_{\rm{em}}p|\left(2|Q^f_{\rm{em}}p|+1\right)
\end{equation}\\
units. Since the minimal electric charge of the Standard Model is $-\frac{1}{3}$, the instanton charge $p$ has to be restricted to\,\cite{Holman:1992ah}
\begin{equation}
    p=3n,~~\text{for some integer $n$}\,,
\end{equation}
and eq.\,\eqref{eq:Zgrav 1 inst, Eguchi Hanson} becomes
\begin{align}
\left.\mathcal{Z}_{\rm{E}}\right|_{\rm{1-inst}}&=\sum_{p=3n}\int\left[\mathcal{D}\mathcal{A}\right]\left[\mathcal{D}\Psi\right]\left[\mathcal{D}h\right]\times\nn\\
&\times e^{-\frac{\pi  p^2}{\alpha}+i\,\frac{p^2}{2} \theta _{\text{em}}+i\frac{\theta_{\rm{grav}}}{16}+\dots}+\mathcal{O}(e^{-M_{\rm{Pl}}^2{\rho}^2})\nn\\
&=\int\left[\mathcal{D}\mathcal{A}\right]\left[\mathcal{D}\Psi\right]\left[\mathcal{D}h\right]\times\nn\\&\hspace{-1.5cm}\times e^{-\frac{9\,\pi }{\alpha}+i\frac{9}{2}  \theta_{\text{em}}+i\frac{\theta_{\rm{grav}}}{16}+\dots}+\mathcal{O}\left(e^{-M_{\rm{Pl}}^2{\rho}^2},\,e^{-\frac{27\pi}{\alpha}}\right)\,,
\end{align}
where, in the following, we will exclude $p=0$ instantons, as they do not violate the chiral charge and therefore do not contribute to the axion potential. Equivalently, since there is no violation of the chiral charge \eqref{eq:Chiral charge}, it follows that no 't Hooft operator is generated for $p=0$. Therefore, the 't Hooft operator, generalizing the QCD one in eq.\,\eqref{eq:'t Hooft operator QCD 1}, associated with $p=3$ AEH instantons will be
\begin{widetext} 
\begin{equation} \label{eq:gravitational 't Hooft operator 1}
\Delta\mathcal{L}_{\rm{AEH}}(x)=\mathcal{K}\left(\frac{9\pi}{\alpha}\right)^{5/2}\int_{m_{\rm{UV }}^{-1}}\frac{d\rho}{\rho^5}\det\left[\prod_{f=\rm{fermions}} \left(\overline{\psi}_{R,f}\psi_{L,f}\right)^{3|Q_{\rm{em}}^f|\left(6|Q_{\rm{em}}^f|+1\right)}\right]e^{-\frac{9\pi}{\alpha}+i\frac{9}{2}  \theta_{\text{em}}+i\frac{\theta_{\rm{grav}}}{16}}+\rm{h.c.}\,,
\end{equation}
\end{widetext}
where $\mathcal{K}$ is an $O(1)$ number, we have used the fact that the chiral charge in eq.\,\eqref{eq:Chiral charge} is violated by $\Delta Q_5=6|Q_{\rm{em}}^f|\left(6|Q_{\rm{em}}^f|+1\right)$ units, that the abelian gauge connection has 5 zero modes (see appendix\,\ref{Eguchi-Hanson metric} for details) and the product in eq.\,\eqref{eq:gravitational 't Hooft operator 1} runs over all the fermions at disposal. In the presence of an axion, the topological angles in eq.\,\eqref{eq:Topological terms} are shifted to
\begin{equation}\label{eq:axion shift}
\begin{gathered}
\theta_{\rm{em}}\rightarrow\theta_{\rm{em}}+\frac{2E a}{v_a}\,,~~~~~
\theta_{\rm{grav}}\rightarrow\theta_{\rm{grav}}+\frac{2G a}{v_a}\,.
\end{gathered}
\end{equation}
Therefore, the exponential in eq.\,\eqref{eq:gravitational 't Hooft operator 1} is modified to 
\begin{equation} \label{eq:gravitational 't Hooft operator 2}
\exp\left(-\frac{9\pi}{\alpha}+i  \theta_{\text{AEH}}+i\mathcal{E}_{\rm{AEH}}\frac{a}{f_a}\right)\,,
\end{equation}
where, for simplicity, we have defined $\,\theta_{\rm{AEH}}\equiv\frac{9}{2}\theta_{\rm{em}}+\frac{\theta_{\rm{grav}}}{16}$ and $\mathcal{E}_{\rm{AEH}}\equiv\frac{9E}{2N}+\frac{G}{16N}$. As in the QCD case, since eq.\,\eqref{eq:gravitational 't Hooft operator 2} is not symmetric under continuous shifts $a(x)/f_a\rightarrow a(x)/f_a+\alpha$, it may be naturally interpreted as a gravitationally induced axion potential. In particular, it should be noted that in eq.\,\eqref{eq:gravitational 't Hooft operator 1} we are integrating up to (see eq.\,(\ref{eq:PlanckMassCoupling}))
\begin{equation}
m_{\rm{UV}}<M_{s}=g_{s}M_{\rm{Pl}}\,,
\end{equation}
because the instanton solution we are considering is a solution of Einstein's equations, and, as commonly believed, GR is not the complete theory of gravity. However, assuming that the UV completion of gravity is weakly coupled, $M_{s}\ll M_{\rm{Pl}}$, \textit{i.e.} that the new degrees of freedom associated with quantum gravity must appear \textit{before} $M_{\rm{Pl}}$, these non-perturbative effects arising from a UV completion of gravity will be weighted by the Euclidean action
\begin{equation}
\sim\exp\left(-\frac{\rm{const}}{g_{s}}\right)\ll\exp\left(-\frac{\rm{9\pi}}{\alpha}\right)\,,
\end{equation}
where the inequality follows if we consider $g_{s}\ll\alpha$; under this assumption,   $M_{s}^4/g_s^2\exp(-9\pi/g_{s})\ll V_{\rm{AEH}}(a)$. We can depict this separation of scales as in fig.\,\ref{Fig:scales}. This allows us to conclude that, under the assumption that the UV completion of GR is weakly coupled, with $g_{s}\ll\alpha$, the gravitationally induced axion potential is dominated by effects of classical GR. 
 That is to say
\begin{align}
\Delta\mathcal{L}_{\rm{AEH}}(x)&=\left.\Delta\mathcal{L}_{\rm{AEH}}(x)\right|_{\rm{GR}}+\mathcal{O}\left(e^{-\frac{\rm{const}}{g_{s}}}\right)\nn\\
&\simeq\left.\Delta\mathcal{L}_{\rm{AEH}}(x)\right|_{\rm{GR}}\,.
\end{align}
\begin{figure*}
    \includegraphics[width=0.95\textwidth]{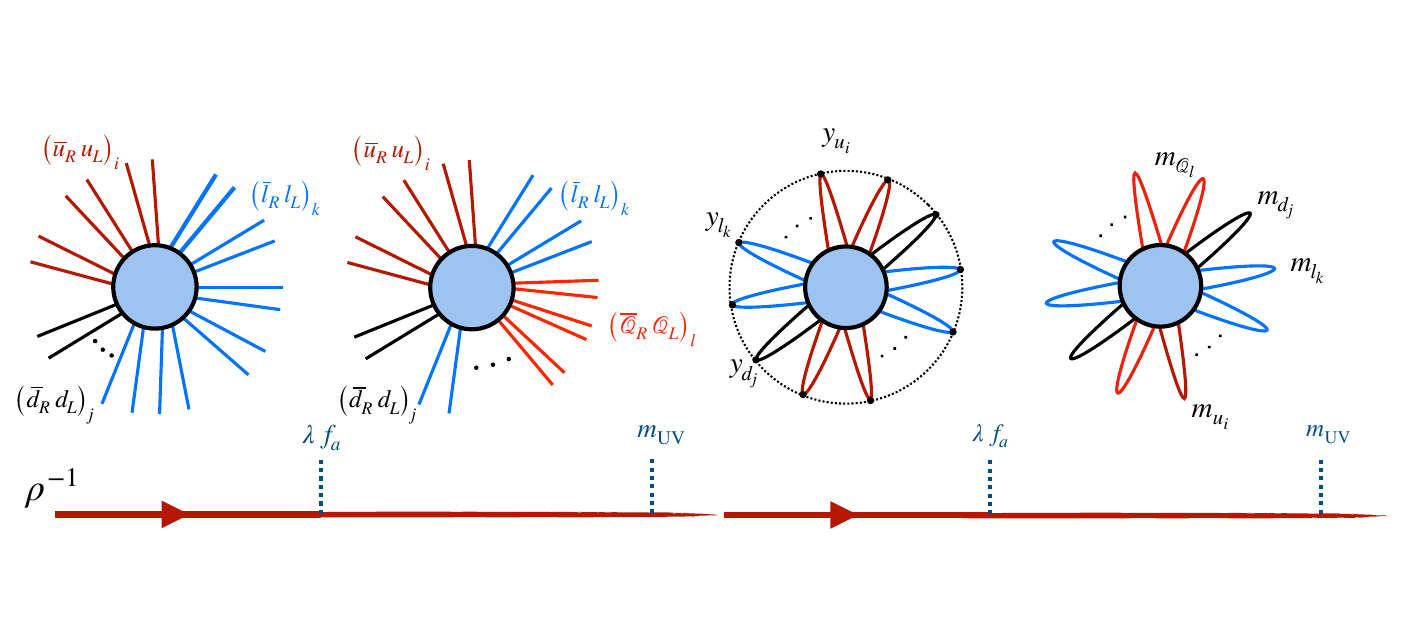}
    \caption{\em Diagrammatic representation of AEH instanton vertex (on the left). AEH instanton contribution to the axion potential (on the right).}
  \label{fig:Master diagrams}
\end{figure*}
In the following $\Delta\mathcal{L}_{\rm{AEH}}(x)=\left.\Delta\mathcal{L}_{\rm{AEH}}(x)\right|_{\rm{GR}}$. Taking as a proxy $\lambda f_a\simeq 10^{12}\rm{GeV}$, where $\lambda$ is some coupling, we can compute numerically the axion potential induced by eq.\,\eqref{eq:gravitational 't Hooft operator 2}. As long as we are at energy scales below $\lambda\,f_a$ we will not be able to probe the onset of the UV completion of the axion model, therefore the product in eq.\,\eqref{eq:gravitational 't Hooft operator 2} will run over Standard Model fermions.
Therefore, omitting the color indices, the integral in eq.\,(\ref{eq:gravitational 't Hooft operator 1}) separates into two distinct contributions, which we write as 
\begin{align}\label{eq:operator3}
\Delta\mathcal{L}_{\rm{AEH}}(x)&=\mathcal{K}\left(\frac{9\pi}{\alpha}\right)^{5/2}\int_{\left(\lambda\, f_a\right)^{-1}}\frac{d\rho}{\rho^5}~\times\nn\\
&\times\det\left[\left\{\left(\overline{l}_Rl_L\right)^{21}\left(\overline{u}_Ru_L\right)^{10}\left(\overline{d}_Rd_L\right)^3\right\}^{N_g}\right]\times\nn\\
&\times e^{-\frac{9\pi}{\alpha}+i\theta_{\rm{AEH}}+i\mathcal{E}_{\rm{AEH}}\frac{a}{f_a}}\nn\\
&+\mathcal{K}\left(\frac{9\pi}{\alpha}\right)^{5/2}\int^{\left(\lambda\, f_a\right)^{-1}}_{m_{\rm{UV}}^{-1}}\frac{d\rho}{\rho^5}~\times\nn\\
&\times\det\Bigg[\left\{\left(\overline{l}_Rl_L\right)^{21}\left(\overline{u}_Ru_L\right)^{10}\left(\overline{d}_Rd_L\right)^3\right\}^{N_g}\times\nn\\
&\times\prod_{\mathcal{Q}}\left(\overline{\mathcal{Q}}_R\mathcal{Q}_L\right)^{3|Q_{\rm{em}}^{\mathcal{Q}}|\left(6|Q_{\rm{em}}^{\mathcal{Q}}|+1\right)}\Bigg]\times\nn\\
&\times e^{-\frac{9\pi}{\alpha}+i\theta_{\rm{AEH}}+i\mathcal{E}_{\rm{AEH}}\frac{a}{f_a}}+\rm{h.c.}\,, 
\end{align}
where $N_g$ is the number of generations of light fermions and we have taken into account, for $\lambda\,f_a<\rho^{-1}<m_{\rm{UV}}$ the onset of the axion model UV completion, \textit{i.e.} the heavy fermions $\mathcal{Q}$, since there are additional heavy degrees of freedom (arising above $m_{\rm{UV}}$) that we have integrated out in our treatment. Diagrammatically, this instanton vertex is represented on the left side of fig.\,\ref{fig:Master diagrams}.\\We now examine how to translate this interaction into a potential for the axion. 
There are two important points to examine.
\begin{itemize}
\item[$\circ$] Let us consider the first of the two integrals in eq.\,(\ref{eq:operator3}), the one bounded below by the length scale 
$(\lambda f_a)^{-1}$, schematically
\begin{equation}\label{eq:first piece}
    \int_{(\lambda f_a)^{-1}}^{M_W^{-1}}\frac{d\rho}{\rho^5}f(\psi_{\rm{light}})\,,
\end{equation}
where $f(\psi_{\rm{light}})$ denotes the 't Hooft operator constructed from the Standard Model fermions, collectively denoted by $\psi_{\rm{light}}$. For concreteness we take the W boson mass $M_W$ as a proxy for the upper integration limit. The fermion legs in $f(\psi_{\rm{light}})$ can be closed using mass insertions and/or Yukawa interactions. Each time a mass insertion is employed (see appendix~\ref{Closing fermion loops with Yukawa} and~\cite{Csaki:2023ziz}), we trade a fermion bilinear $\overline\psi\psi$ for $m_\psi\,\rho$ --- $m_f$ being the mass of the corresponding fermion $\psi_f$ --- thereby positive powers of $\rho$ are introduced into the integrand. Instead, a Yukawa interaction replaces the bilinear  $\overline\psi\psi$ with the Yukawa coupling $y_\psi$ without altering the integrand's $\rho$-dependence. Depending on how we close the light fermion modes $\psi_{\rm{light}}$, \eqref{eq:first piece} gives rise to a sum of different contributions to the axion potential $V_{\rm{AEH}}(a)$, all of which will be suppressed by the smallness of the Yukawa couplings and masses of Standard Model leptons. Among these, we find numerically that the leading contribution --- given that $\lambda f_a\gg m_{SM}$, where $m_{SM}$ is any Standard Model mass --- comes from closing $\psi_{\rm{light}}$ with Yukawa interactions. In this case, no additional powers of $\rho$ are introduced and the potential scales as $(\lambda f_a)^4$. 
In the right panel of fig.\,\eqref{fig:Master diagrams} we show the numerically leading contribution obtained from \eqref{eq:first piece}, where $\psi_{\rm{light}}$ are closed with Yukawa interactions\footnote{In principle the zero modes could be closed using other Standard Model interactions, such as gauge interactions and Higgs boson exchanges. However these contributions are further suppressed by additional loop factors and powers of the corresponding gauge couplings~\cite{DINE1986109}.}.
\item[$\circ$] Let us now consider the second integral, defined over length scales in the range 
$\lambda\,f_a<\rho^{-1}<m_{\rm{UV}}$, schematically
\begin{equation}\label{eq:second piece}
    \int_{m_{\rm{UV}}^{-1}}^{(\lambda f_a)^{-1}}\frac{d\rho}{\rho^5}f(\psi_{\rm{light}},\psi_{\rm{heavy}})\,,
\end{equation}
where $f(\psi_{\rm{light}},\psi_{\rm{heavy}})$ is the 't Hooft operator constructed from the Standard Model fermions and fermions of the completion of the axion model, the latter collectively denoted by $\psi_{\rm{heavy}}$. This vertex can be reliably used \textit{only} well below $m_{\rm{UV}}$, while contributions corresponding to the upper integration limit will be of the order of $\mathcal{O}(e^{-\frac{\rm{const}}{g_s}})$. Therefore, the contribution to the axion potential from the AEH solution comes from closing the fermion legs with mass insertions, resulting in an integral that exhibits the dominant behavior near the scale
$\left(\lambda\,f_a\right)^{-1}$.
This is shown, in the corresponding range of lengths, in the right panel of fig.\,\ref{fig:Master diagrams}.
\end{itemize}
All in all, we obtain the axion potential contribution (see appendix~\ref{Closing fermion loops with Yukawa}), up to $O(1)$ factors
\begin{align}
V_{\rm{AEH}}(a)&=\lambda^3\hbar^{17N_g}\left(\frac{9\pi}{\alpha}\right)^{5/2}\frac{\left(\lambda\,f_a\right)^4}{(\pi^2)^{17N_g}}\times\nn\\
&\times\prod_{i=1}^{N_g}\left(y_l^{21}\, y_u^{10} \,y_d^3\right)_i e^{-\frac{9\pi}{\alpha}+i\theta_{\rm{AEH}}+i\mathcal{E}_{\rm{AEH}}\frac{a}{f_a}}\nn\\
&+\lambda^3\left(\frac{9\pi}{\alpha}\right)^{5/2}\left(\lambda f_a\right)^{4-34N_g-10N_{\mathcal{Q}}}\times\nn\\
&\times \prod_{i=1}^{N_g}\left(m_l^{21}m_u^{10}m_d^3\right)_i\prod_{\mathcal{Q}=1}^{N_{\mathcal{Q}}}M_{\mathcal{Q}}^{10}\times\nn\\
&\times e^{-\frac{9\pi}{\alpha}+i\theta_{\rm{AEH}}+i\tilde{\mathcal{E}}_{\rm{AEH}}\frac{a}{f_a}}\,+\rm{h.c.}\,,
\end{align}
where, for dimensional reasons, we have one overall factor of $\lambda^3$, and, for concreteness, we consider the case in which $Q_{\rm{em}}^{\mathcal{Q}}=2/3$.
We find numerically that, since we want to enhance as much as possible this effect, taking $\lambda=O(1)$, $f_a=10^{12}\,\rm{GeV}$ and $\alpha=10^{-1}$, because of loop and mass suppressions (see appendix\,\ref{Closing fermion loops with Yukawa})
\begin{align}
V_{\rm{AEH}}(a)&\ll\underbrace{\lambda^3\left(\frac{9\pi}{\alpha}\right)^{5/2}e^{-\frac{9\pi}{\alpha}}\left({\lambda}f_a\right)^4}_{\simeq10^{-70} \,\rm{GeV}^4}\,\times\nn\\&\times\cos\left(\theta_{\rm{AEH}}+\mathcal{E}_{\rm{AEH}}\frac{a}{f_a}\right)\,.\label{eq:MainAEH}
\end{align}
We may try to enhance this potential closing the fermion legs via Yukawa couplings associated with scalars that are not the Standard Model Higgs\,\cite{FLYNN1987731}. Nevertheless, we find that (taking the Yukawa couplings to be $O(1)$ numbers) this effect is still negligible compared to the QCD contribution.
\subsection{Comment on the colored case}
We notice that we could have taken into account in the Euclidean generating functional $\mathcal{Z}_{\rm{E}}$ in eq.\,\eqref{eq:Zgrav 1 inst} also the so-called CEH instantons\,\cite{Chen:2021jcb}. CEH instantons have Euclidean action
\begin{equation}
S_{\rm{E}}\left[\overline{g}_{\rm{CEH}}\right]=\frac{4\pi^2 }{g_S^2}\times3\,+i\frac{3}{2}\,\overline{\theta}+i\frac{\theta_{\rm{grav}}}{16}\,.
\end{equation}
Using similar arguments for the abelian case, one gets the 't Hooft vertex
\begin{align}\label{eq:Colored Eguchi-Hanson}
\Delta\mathcal{L}_{\rm{CEH}}(x)&=\tilde{\mathcal{K}}\left(\frac{3\pi}{\alpha_S}\right)^8\int_{\left(\lambda\,f_a\right)^{-1}}^{\Lambda^{-1}}e^{-S_{\rm{E}}\left[\overline{g}_{\rm{CEH}}\right]}\,\frac{d\rho}{\rho^5}\times\nn\\
&\times\prod_{\rm{QCD~quarks}}\left(\overline{q}_L\,q_R\right)~+\nn\\
&+\tilde{\mathcal{K}}\left(\frac{3\pi}{\alpha_S}\right)^8\int_{m_{\rm{UV}}^{-1}}^{\left(\lambda\,f_a\right)^{-1}}e^{-S_{\rm{E}}\left[\overline{g}_{\rm{CEH}}\right]}\,\frac{d\rho}{\rho^5}\times\nn\\
&\times\prod_{\rm{QCD~quarks}}\left(\overline{q}_L\,q_R\right)~\prod_{\mathcal{Q}}\left(\overline{\mathcal{Q}}_L\,\mathcal{Q}_R\right)^{f(R_{\mathcal{Q}})}\nn\\
&+\rm{h.c.}\,,
\end{align}
where $\tilde{\mathcal{K}}$ is a $O(1)$ number, $g_S$ is the strong coupling, $f(R_\mathcal{Q})$ is the number of zero-modes associated to the color representation $R_\mathcal{Q}$ under which $\mathcal{Q}$ transforms\,\cite{Bianchi_1995}, $\Lambda$ is an infrared scale and, shifting $\overline\theta\to\overline\theta+\frac{a}{f_a}$, the exponential in eq.\,\eqref{eq:Colored Eguchi-Hanson} is modified to
\begin{equation}
\exp\left(\frac{4\pi^2}{g_S^2}\times3+i{\theta}_{\rm{CEH}}+i\mathcal{A}_{\rm{CEH}}\frac{a}{f_a}\right)\,,
\end{equation} in which ${\theta}_{\rm{CEH}}=\frac{3}{2}\overline{\theta}+\frac{\theta_{\rm{grav}}}{16}$ and $\mathcal{A}_{\rm{CEH}}=\frac{3}{2}+\frac{G}{16N}$. All the calculations involving the 't Hooft operator in eq.\,\eqref{eq:'t Hooft operator QCD 1}, namely closing external fermion legs to get a potential, are valid as long as the coupling associated with the gauge connection is \textit{perturbative}, since in the perturbative domain one-instanton contributions saturate the axion potential. In other words, closing the legs of the 't Hooft vertices in the strong coupling regime is unlikely to yield reliable results. Supporting this argument is the fact that the mass of the $\eta'$ is not governed by large instanton effects, but rather by the confinement dynamics\,\cite{Csaki:2023yas}. In light of these observations, in the second contribution in eq.\,\eqref{eq:Colored Eguchi-Hanson}, \textit{i.e.} the one in which $\lambda\,f_a<\rho^{-1}<m_{\rm{UV}}$, the coupling is perturbative. Consequently, using 't Hooft operators to calculate the axion potential can yield meaningful results. For the instanton calculation to produce a reliable result, in the integration region $(\lambda\,f_a)^{-1}<\rho <\Lambda^{-1}$ where $\alpha_S(\Lambda)=O(1)$, the legs of the vertex in eq.\,\eqref{eq:Colored Eguchi-Hanson} can be closed only through interactions that force the lower limit of integration to dominate, since for $\rho^{-1}=\lambda\,f_a$, $g_S$ is perturbative. Therefore, to avoid the large instanton region, we must not introduce positive powers of $\rho$, differently from the analysis in\,\cite{Chen:2021jcb}. The only way to do avoid the strong coupling regime is by closing the legs with Yukawa interactions. Because of the perturbativity of $\alpha_S$ and loop suppressions, we get, for $\lambda=O(1)$ and $\alpha_S=10^{-2}$
\begin{equation}
V_{\rm{CEH}}(a)\lsim \mathcal{O}(10^{-200})\,V_{\rm{QCD}}(a)\,,
\end{equation}
where
\begin{equation*}
V_{\rm{QCD}}(a)=-m_{\pi}^2 f_{\pi}^2\sqrt{1-\frac{4m_u m_d}{(m_u+m_d)^2}\sin^2\left(\frac{a}{2f_a}\right)}\,.
\end{equation*}\\
This therefore shows that CEH instantons provide a subleading contribution to the axion potential compared to the abelian case.
\subsection{Gravitational potential enhancement in \textit{Nnatualness}}
The only beyond the Standard Model scenario in which these gravitational effects can be significantly enhanced involves introducing multiple copies of the Standard Model to address the Higgs boson mass hierarchy problem\,\cite{Nnaturalness}. Having $N$ copies of the SM, naively implies that each copy has its own theta angle $\theta_i$, however, if the $S_N$ symmetry distinguishing among the different copies is only softly broken by Higgs mass terms (\textit{i.e.} these copies differ only for the mass term in the Higgs potential), all these angles would be equal. Therefore, a shared axion would be able to set to zero all the $\theta_i$'s at the same time. Moreover, if one naively assumes the presence of an axion in each sector, this would elevate the number of light degrees of freedom, consequently worsening the constraints on $N_{\rm{eff}}$. If a shared axion is the solution to the strong CP problem, then the number of copies is bounded to $N<10^{10}$\,\cite{Nnaturalness}. In this case, the 't Hooft operator will be
\begin{widetext} 
\begin{equation}
\Delta\mathcal{L}_{\rm{AEH}}(x)=\sum_{\rm{SM}=\rm{us}}^{10^{10}}\left.\left\{\left(\frac{9\pi}{\alpha}\right)^{5/2}\int_{m_{\rm{UV }}^{-1}}\frac{d\rho}{\rho^5}\det\left[\prod_{f=\rm{fermions}} \left(\overline{\psi}_{R,f}\psi_{L,f}\right)^{3|Q_{\rm{em}}^f|\left(6|Q_{\rm{em}}^f|+1\right)}\right]e^{-S_{\rm{E}}}+\rm{h.c.}\right\}\right|_{\rm{SM}}\,,
\end{equation}
\end{widetext}
where, as usual 
\begin{equation*}
S_{\rm{E}}=e^{-\frac{9\pi}{\alpha}+i\theta_{\rm{AEH}}+i\mathcal{E}_{\rm{AEH}}\frac{a}{f_a}}\,.
\end{equation*}
Taking the Yukawa couplings to be $O(1)$ numbers, one gets that this contribution is still significantly small compared to the QCD contribution. Therefore, this gravitational contribution to the axion potential is still exceedingly small with respect to the axion potential generated by confinement dynamics in the strong interactions.

\section{Conclusions}\label{sec:conclusions}
In this work, we analyzed the role of self-dual ALE solutions of the vacuum Euclidean Einstein equations in the Euclidean gravitational path integral. Our results indicate that low-energy physics potentially sensitive to these contributions—particularly the QCD axion—remains protected. This protection persists even across a range of beyond Standard Model scenarios, which might otherwise enhance these effects. Adopting a maximally inclusive definition of the path integral (as discussed in section\,\ref{sec:Vermi}), we derived an upper bound on the gravitational contribution to the axion potential. Our key findings are summarized as follows:
\begin{itemize}

\item[$\circ$] \textbf{AEH instantons and weakly coupled string theory}: Assuming a weakly coupled string theory as the UV completion of gravity, we argued that AEH instantons saturate the gravitational contribution to the axion potential.

\item[$\circ$] 
\textbf{Zero modes and loop suppressions}: Following ref.\,\cite{Franchetti_2018},  we identified zero modes previously overlooked in refs.\,\cite{Holman:1992ah, Rey:1992qt}. These additional zero modes, along with loop and mass suppressions, significantly reduce the AEH instanton contribution to the axion potential, see eq.\,(\ref{eq:MainAEH}). Specifically, the abelian contribution remains well below the current experimental bounds on the neutron electric dipole moment, independent of the axion model's UV completion.

\item[$\circ$] 
\textbf{CEH instantons and PQ protection}:
By including CEH instantons in the sum over ALE manifolds in the gravitational path integral, see eq.\,\eqref{eq:Zgrav 1 inst}, we showed that—contrary to the results of ref.\,\cite{Chen:2021jcb}---these do not spoil the PQ solution. Instead, their contribution is subleading compared to the abelian case.

\item[$\circ$] 
\textbf{PQ quality problem in beyond the Standard Model scenarios}:
Our analysis demonstrates that this non-perturbative contribution to the axion potential does not introduce a PQ quality problem, even in extended beyond the Standard Model scenarios. For instance, in the model discussed in ref.\,\cite{Nnaturalness}, where multiple Standard Model copies are coupled to a single $U(1)_{\textrm{PQ}}$, the PQ symmetry remains protected.
\end{itemize}

In conclusion, our findings suggest that gravitational effects do not pose a threat to the robustness of the QCD axion solution. This strengthens the viability of axion-based solutions to the strong CP problem across a wide range of UV completions and beyond the Standard Model frameworks.

\acknowledgments

The authors thank R. Contino, T. Cohen, R. T. D'Agnolo, A. Hebecker and D. E. Kaplan for enlightening discussions.
This work is partially
supported by ICSC - Centro Nazionale di Ricerca in
High Performance Computing, Big Data and Quantum
Computing, funded by European Union-NextGenerationEU.

\appendix

\section{Eguchi-Hanson manifold}\label{Eguchi-Hanson metric}
\begin{figure}[h!]
    \includegraphics[width=0.43\textwidth]{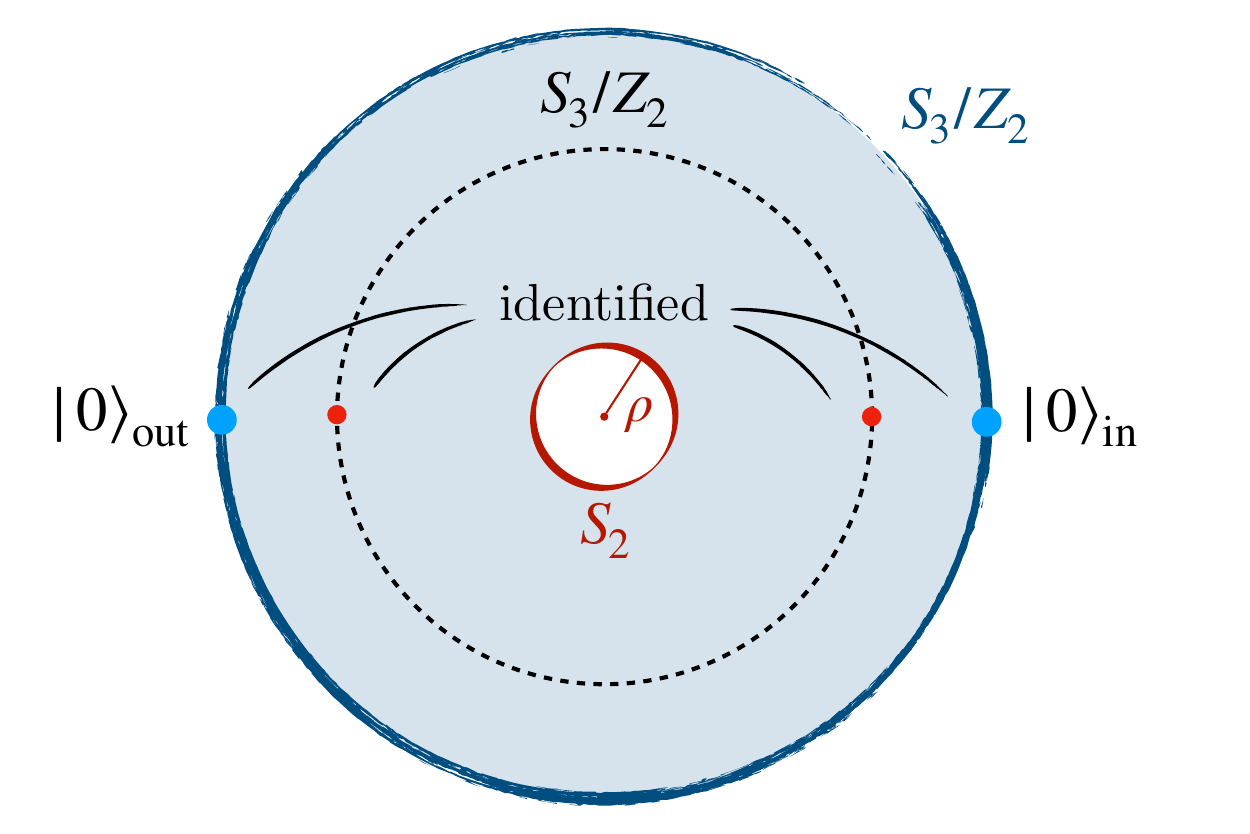}
    \caption{\em Schematical representation of $\mathcal{M}_{\rm{(A)EH}}$ spacetime.}
    \label{fig:Schematical representation of AEH spacetime}
\end{figure}

\begin{figure}[h!]
    \includegraphics[width=0.43\textwidth]{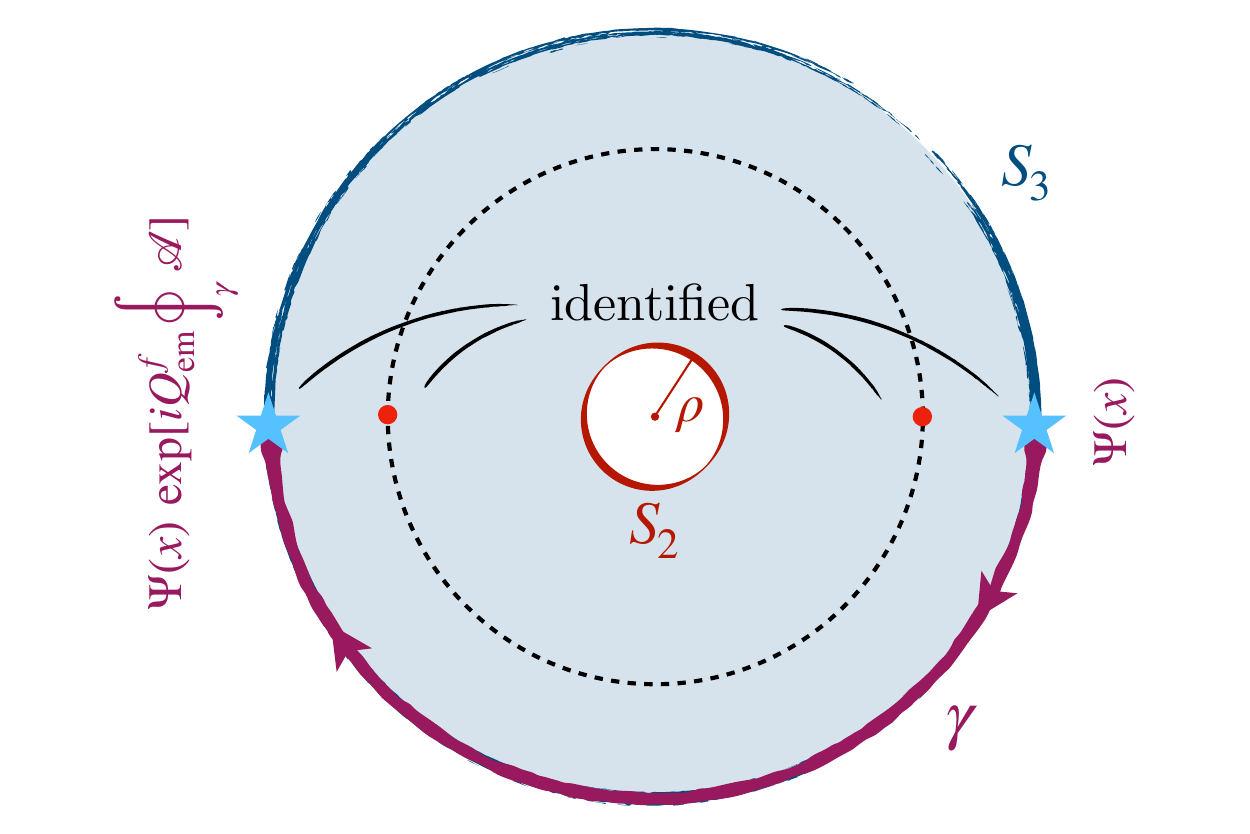}
    \caption{\em Propagation of the fermion $\Psi(x)$ along the loop $\gamma$.}
    \label{fig:propagation}
\end{figure}

Eguchi and Hanson discovered a non-trivial spherically symmetric solution of Einstein's equations in 4-dimensional Euclidean spacetime\,\cite{Eguchi:1976db,Eguchi:1978gw}, the so-called \textit{Eguchi-Hanson instanton}.
One would expect that the relevant instanton-like metrics would be those whose 2-form curvature $R_{ab}$ is (anti)-self-dual, are localized in Euclidean spacetime and free of singularities. In fact, solutions have been found which have the interesting property that the metric approachs a flat metric at infinity, i.e. a vacuum configuration (as required for a tunneling interpretation). These solutions are called ``asymptotically locally Euclidean'' metrics, because as we shall see, at infinity, they are locally Euclidean. Apart for this solutions, the obvious metrics that seem to incarnate these features (self-dual, localized and free of singularities) are the standard solutions of black hole physics, which however in Euclidean spacetime are not asymptotically flat. The instanton metric is
\begin{align}\label{eq:Eguchi Hanson metric}
    ds^2_{\rm EH}&=\frac{\text{dr}^2}{1-\frac{\rho^4 }{r^4}}+\frac{r^2}{4}  (\sigma_x^2+\sigma_y^2)+\frac{r^2}{4}  \sigma_z^2 \left(1-\frac{\rho^4}{r^4}\right)\nn\\
    &=\frac{dr^2}{1-\frac{\rho^4}{r^4}}+\frac{r^2}{4}\left[d\theta^2+\sin^2\theta\,d\phi^2\right]\nn\\
    &+\frac{r^2}{4}\left(1-\frac{\rho^4}{r^4}\right)(d\psi+\cos\theta\, d\phi)^2\,,~\,r\geq\rho\,,
\end{align}
where we introduced the 1-forms
\begin{equation*}
    \begin{aligned}
        \sigma_1&=\sin\psi\,d\theta-\cos\psi\sin\theta d\phi\,,\\
        \sigma_2&=\cos\psi d\theta+\sin\psi\sin\theta d\phi\,,\\
        \sigma_3&=d\psi+\cos\theta d\phi\,,
    \end{aligned}
\end{equation*}
$(\phi,\theta,\psi)$ are the Euler angles on the $S^3$ sphere with ranges
\begin{equation*}
        0\leq\theta<\pi\,,~
        0\leq\phi<2\pi\,,~
        0\leq\psi<4\pi\,,
\end{equation*}
and $\rho$ is the instanton core size (see Fig.~\ref{fig:Schematical representation of AEH spacetime}). That is, in matrix form
\begin{align*}
g_{\mu\nu}^{\rm{EH}}=\left(
\begin{array}{cccc}
 \frac{r^2}{4} & 0 & 0 & 0 \\
 0 & -\frac{\rho ^4 \cos 2 \theta +\rho ^4-2 r^4}{8 r^2} & \frac{\left(r^4-\rho ^4\right)\cos \theta  }{4 r^2} & 0 \\
 0 & \frac{ \left(r^4-\rho ^4\right)\cos \theta }{4 r^2} & \frac{r^4-\rho ^4}{4 r^2} & 0 \\
 0 & 0 & 0 & \frac{1}{1-\frac{\rho ^4}{r^4}} \\
\end{array}
\right)\,.
\end{align*}
We have the vierblein
\begin{equation}
\begin{gathered}
ds^2_{\rm{EH}}=\sum_{a=1}^4\sum_{b=1}^4 e^a e^b \eta_{ab}\,,~e^a=e^a_\mu dx^\mu\\
e^a=\left(\frac{r}{2}\sigma_x,\frac{r}{2}\sigma_y,\frac{u\,r}{2}\sigma_z,-\frac{dr}{u}\right)\,,~u=\sqrt{1-\frac{\rho^4}{r^4}}\,,
\end{gathered}
\end{equation}
where in the Euclidean $\eta_{ab}=\eta^{ab}={\rm{diag}}(1,1,1,1)$. We have the wedge products
\begin{align*}
e^1\wedge e^2&=\frac{r^2}{4}\sin\theta\,d\theta\wedge d\phi\,,\\
e^1\wedge e^3&=\frac{r^2}{4}\,u\big(-\cos\psi\sin\theta\,d\phi\wedge d\psi\\
&+\sin\psi\left[d\theta\wedge d\psi+\cos\theta d\theta\wedge d\phi\right]\big)\,,\\
e^1\wedge e^4&=-\frac{r}{2u}\big( \sin \psi d\theta\wedge dr - \sin \theta \cos \psi d\phi\wedge dr\big)\,,\\
e^2\wedge e^3&=\frac{u\,r^2}{4}\big(\sin\theta\sin\psi\,d\phi\wedge d\psi\\
&+\cos\psi\,\left[d\theta\wedge d\psi+\cos\theta\,d\theta\wedge d\phi\right]\big)\,,\\
e^2\wedge e^4&=-\frac{r}{2u}\big(\cos\psi\,d\theta\wedge dr+\sin\theta\sin\psi\,d\phi\wedge dr\big)\,,\\
e^3\wedge e^4&=-\frac{r}{2}\big(d\psi\wedge dr+\cos\theta\,d\phi\wedge dr\big)\,.
\end{align*}
We have the curvature 2-form
\begin{align*}
    \tensor{R}{^a_b}&=\tensor{R}{^a_b}=\frac{1}{2}\tensor{R}{^a_{b\mu\nu}}dx^\mu\wedge dx^\nu\\
    &\equiv d\tensor{\omega}{^a_b}+\tensor{\omega}{^a_c}\wedge\tensor{\omega}{^c_b}\,,
\end{align*}
where $\tensor{\omega}{^a_b}$ is the spin-connection 1-form\,\cite{Birrell:1982ix,Ortin:2015hya}, and we find the self-dual curvature (for the flat indices $a,\,b$ we are taking $\varepsilon_{1234}=+1$), $R_{mn}=+\frac{1}{2}\varepsilon_{mnab}R_{ab}$,
\begin{align*}
    R_{14}&=R_{23}=-\frac{2\rho^4}{r^6}\left(e^1\wedge e^4+e^2\wedge e^3\right)\,,\\
    R_{24}&=-R_{13}=-\frac{2\rho^4}{r^6}\left(e^2\wedge e^4+e^1\wedge e^3\right)\,,\\
    R_{34}&=R_{12}=+\frac{4\rho^4}{r^6}\left(e^3\wedge e^4+e^1\wedge e^2\right)\,.
\end{align*} 
To remove the coordinate singularity at $r=\rho$, we have to identify antipodal points of $r=a>\rho$ spherical surfaces. That is to say
\begin{equation}
    \begin{aligned}
        0&\leq\theta<\pi,\\
        0&\leq\phi<2\pi,\\
        0&\leq\psi<2\pi\,.
    \end{aligned}
\end{equation}
This is an explicit example of a metric whose topology is asimptotically \textit{locally} Euclidean, but not \textit{globally} Euclidean (i.e., not $S^3$). We have the Pontryagin invariant
\begin{align*}
    p&=-\frac{1}{16\pi^2}\int_{\mathcal{M}_{\rm{EH}}}d^4x\,\sqrt{g}R_{\mu\nu\rho\sigma}\tilde{R}^{\mu\nu\rho\sigma}\nn\\
    &=\frac{1}{16\pi^2}\int_{\mathcal{M}_{\rm{EH}}}d^4x\,\sqrt{g}\,\frac{384\rho^8}{r^{12}}=3\,.
    \end{align*}
The EH metric \eqref{eq:Eguchi Hanson metric} supports the self-dual $U(1)$ field 
\begin{equation}\label{eq:gauge field eguchi hanson}
    \mathcal{A}=\frac{p\rho^2}{r^2}\sigma_z,
\end{equation}
which is to say
\begin{equation}
    \mathcal{A}_r=\mathcal{A}_\theta=0,~~\mathcal{A}_\psi=\frac{p\rho^2}{r^2},~~\mathcal{A}_\phi=\frac{p\rho^2}{r^2}\cos\theta\,.
\end{equation}
This gauge field connection has $5$ collective coordinates: $4$ related to the instanton center $x_0$, 1 related to the integration constant $\rho$. We have the field-strength 
\begin{align}
\mathcal{F}&=d\mathcal{A}=\partial_\mu \mathcal{A}_\nu\,dx^\mu\wedge dx^\nu\nn\\
&=-\frac{2p\rho^2\cos\theta}{r^3}\,dr\wedge d\phi\nn\\
&-\frac{2p\rho^2}{r^3}\,dr\wedge d\psi-\frac{p\rho^2\sin\theta}{r^2}\,d\theta\wedge d\phi\nn\\
&=-\frac{4p\rho^2}{r^4}\big(e^1\wedge e^2-e^3\wedge e^4\big)\,.
\end{align}
That is, in matrix form
\begin{align*}
\mathcal{F}_{\mu\nu}=\left(
\begin{array}{cccc}
 0 & -\frac{p \rho ^2 \sin \theta }{r^2} & 0 & 0 \\
 \frac{p \rho ^2 \sin \theta }{r^2} & 0 & 0 & \frac{2 p \rho ^2 \cos \theta }{r^3} \\
 0 & 0 & 0 & \frac{2 p \rho ^2}{r^3} \\
 0 & -\frac{2 p \rho ^2 \cos \theta }{r^3} & -\frac{2 p \rho ^2}{r^3} & 0 \\
\end{array}
\right)
\end{align*}
which is anti-self-dual, that is
\begin{align*}
\tilde{\mathcal{F}}^{\mu\nu}=\frac{1}{2\sqrt{g}}\,\varepsilon^{\mu\nu\rho\sigma}\mathcal{F}_{\rho\sigma}=-\mathcal{F}^{\mu\nu}\,,~~\varepsilon^{0123}=1\,.
\end{align*}
Indeed, since (anti)self-dual Maxwell fields have vanishing energy-momentum tensor\footnote{One can easily verify that, using $\mathcal{F}_{12}=\mathcal{F}_{43}=-\frac{4p\rho^2}{r^4}$, obtains $T_{ab}=-\mathcal{F}_{ac}\mathcal{F}_{bc}+\frac{1}{4}\eta_{ab}\mathcal{F}_{cd}\mathcal{F}_{cd}=0$.}, the Einstein equations are undisturbed and, decorating the EH spacetime in eq.\,\eqref{eq:Eguchi Hanson metric} with the gauge field in eq.\,\eqref{eq:gauge field eguchi hanson}, we have an automatic solution of the Einstein-Maxwell equations\,\cite{EGUCHI1980213}. This is what we call an AEH spacetime. We have the action\,\cite{Xiao_2004}
\begin{align}
S_{\rm{AEH}}&=
    -\frac{1}{16\pi G_N}
    \int_{{M}} d^4x\,\sqrt{g}\,R
    -\frac{1}{8\pi G_N}\int_{\partial M}
    d^3\vec{x}
    \,\sqrt{k}\,[K]
    \,\nn\\&
    +\frac{1}{4e^2}\int_{\mathcal{M}_{\rm{AEH}}}d^4x\,\sqrt{g}\,\mathcal{F}_{\mu\nu}\mathcal{F}^{\mu\nu}\nn\\
&+i\int_{\mathcal{M}_{\rm{AEH}}}d^4x\,\sqrt{g}\,\frac{\theta_{\rm{em}}}{32\pi^2}{\mathcal{F}}_{\mu\nu}\tilde{\mathcal{F}}^{\mu\nu}\\&-i\int_{\mathcal{M}_{\rm{AEH}}}d^4x\,\sqrt{g}\,\frac{\theta_{\rm{grav}}}{768\pi^2}R_{\mu\nu\rho\sigma}\tilde{R}^{\mu\nu\rho\sigma}\nn\\
&=\frac{\pi  p^2}{\alpha}-i\,\frac{p^2}{2} \theta _{\text{em}}-i\frac{\theta_{\rm{grav}}}{16}\,.
\end{align}
Note that this action is independent of Newton’s constant and rather similar to that of the Yang-Mills instanton except for a factor of 2. This can be traced back to the identification of antipodal points in EH space, which becomes a half of the (asymptotic) Euclidean space. We now ask whether the AEH space actually admits the existence of spin structures (Fermi fields)\,\cite{HAWKINGPOPE}. Consider a path $\gamma$ at constant $r=a > \rho$ with  $0<\psi<2\pi$ at fixed $(\theta,\phi)$. Since at constant $r >\rho$, $\mathcal{M}_{\rm{EH}} \sim S_3/Z_2$, this path is indeed a loop (see Fig.~\ref{fig:propagation}). Furthermore this loop cannot be contracted to a point, since the loop encloses the $S_2$ sphere at $r = \rho$. Propagating a fermion field $\Psi(x)$ with charge $Q^f_{\rm{em}}$ along $\gamma$, one will obtain a new field $\Psi'(x)$ which is related to the original field by the abelian transformation
\begin{equation}
\Psi'(x)=e^{iQ^f_{\rm{em}}\,\oint_\gamma \mathcal{A}}~\Psi(x)\,.
\end{equation}
However, from the Stokes' theorem, in the limit $a\rightarrow\infty$,
\begin{align*}
\oint_\gamma \mathcal{A}&=\int_{\mathcal{M}_\gamma}\mathcal{F}\\
&=-\int_{\mathcal{M}_\gamma}\frac{4p\rho^2}{r^4}\big(e^1\wedge e^2-e^3\wedge e^4\big)\\
&=\int_{\mathcal{M}_\gamma}\frac{4p\rho^2}{r^4}\frac{r}{2}d\psi\wedge dr=-2\pi p\,,
\end{align*}
where $\mathcal{M}_\gamma$ is the manifold lying over the loop contour $\gamma$. For the fermion $\Psi(x)$ to be consistently defined over the manifold, since antipodal points are identified, we have to impose the Dirac string quantization condition\,\cite{HAWKINGPOPE}
\begin{equation}\label{eq:monopole quantization}
Q_{\rm{em}}^f\,p=\rm{integer}\,.
\end{equation}
\section{Computation of Dirac zero-modes on AEH manifold}\label{Computation of Dirac zero-modes on Eguchi-Hanson manifold}To make the analysis parallel to the Yang-Mills case and to give a precise estimate of all the loop factors we have to find explicitly the zero-modes of the Dirac operator $i\slashed {\mathcal{D}}$ in the AEH background (see\,\cite{C_N_Pope_1981,Franchetti_2018} for a review). In this appendix we find the zero-modes of the Dirac operator for 4-manifolds with isometry group SO(3) and a self-dual Riemann tensor. Introducing the radial functions $f(r),\,a(r),\,b(r),\,c(r)$, the metrics take the form
\begin{equation}\label{eq:General Metrics}
    ds^2=f^2dr^2+a^2\sigma_1^2+b^2\sigma_2^2+c^2\sigma_3^2\,.
\end{equation}
We furthermore introduce the vierblein 
\begin{equation*}
    e^1=a\sigma_1,~~e^2=b\sigma_2,~~e^3=c\sigma_3,~~e^4=-fdr\,.
\end{equation*}
By definition, the inverse vierblein is given by
\begin{equation*}
    E{^\mu_a}=\eta_{ab}g^{\mu\nu}{e}{^b_\nu}~\Leftrightarrow~E^{\mu}_ae_\mu^b=\delta_a^b\,,
\end{equation*}
from which, inverting the $4\times4$ matrix $e_\mu^a$, we straightforwardly find
\begin{equation*}
    E_1=\frac{1}{a}X_1,~~E_2=\frac{1}{b}X_2,~~E_3=\frac{1}{c}X_3,~~E_4=-\frac{1}{f}\partial_r\,,
\end{equation*}
where
\begin{equation*}
    \begin{aligned}
        X_1&=\sin\psi\,\partial_\theta+\cos\psi\,(\cot\theta\,\partial_\psi-\frac{1}{\sin\psi}\partial_\phi)\,,\\
        X_2&=\cos\psi\,\partial_\theta-\sin\psi\,(\cot\theta\,\partial_\psi-\frac{1}{\sin\theta}\partial_\phi)\,,\\
        X_3&=\partial_\psi\,.
    \end{aligned}
\end{equation*}
For many calculations it is convenient to use a proper radial coordinate $R$ defined via
\begin{equation*}
    dR\equiv f\,dr\Leftrightarrow E_4=-\partial_R\,.
\end{equation*}
We are interested in the general form of the Dirac operator in metrics like the one in eq.\,\eqref{eq:General Metrics} and coupled to a spherically symmetric abelian $U(1)$ gauge connection with self-dual field strenght: indeed for (anti)-self-dual field strenght the energy momentum tensor identically vanishes. The gauge potential can then be written as 
\begin{equation*}
    \mathcal{A}=\mathcal{A}_1\sigma_1+\mathcal{A}_2\sigma_2+\mathcal{A}_3\sigma_3\,,
\end{equation*}
where $\mathcal{A}_1$, $\mathcal{A}_2$, $\mathcal{A}_3$ are functions of $R$ only. The Dirac operator in the presence of the connection $\mathcal{A}$ is 
\begin{align*}
    \gamma^\mu \mathcal{D}_{\mu,\mathcal{A}}&=\slashed{\mathcal{D}}_\mathcal{A}=\gamma^a E^\mu_a\left(\partial_\mu-\frac{1}{8}\tensor{\omega}{_\mu^{bc}}\left[\gamma^b,\gamma^c\right]+iQ^f_{\rm{em}}\mathcal{A}_\mu\right)\nn\\
    &=\gamma^a\bigg(E_a-\frac{1}{8}\tensor{\omega}{_a^{bc}}\left[\gamma^b,\gamma^c\right]+iQ^f_{\rm{em}}\mathcal{A}(E_a)\bigg)\,,
\end{align*}
where $Q^f_{\rm{em}}$ is the charge of the fermion under the U(1) connection $\mathcal{A}$, $\mathcal{A}(E_a)=E^\mu_a\mathcal{A}_\mu$ and we introduced the Euclidean $\gamma$ matrices
\begin{equation}
    \left\{\gamma_a,\gamma_b\right\}=-2\delta_{ab}\,,~~a,b=1,\dots,4\,,
\end{equation}
where
\begin{equation*}
    \gamma^i=\begin{pmatrix}
        0&\tau_i\\
        -\tau_i&0
    \end{pmatrix},~~~\gamma^4=-i\begin{pmatrix}
        0&\mathbf{1}\\
        \mathbf{1}&0
    \end{pmatrix}\,,
\end{equation*}
and we have the Pauli matrices $[\tau_i,\tau_j]=2i\epsilon_{ijk}\tau_k\,,i=1,2,3$.
The self-duality of the Riemann tensor implies that the spin connection 2-form $\omega_{ab}$ is
\begin{equation}\label{eq:General Spin Connection}
    \begin{aligned}
    \omega_{14}&=-\dot{a}\sigma_1,&\omega_{24}&=-\dot{b}\sigma_2,&\omega_{34}&=-\dot{c}\sigma_3\,,\\
        \omega_{23}&=-A\sigma_1,&\omega_{31}&=-B\sigma_2,&\omega_{12}&=-C\sigma_3\,,
    \end{aligned}
\end{equation}
where
\begin{equation*}
    A=\frac{b^2+c^2-a^2}{2bc},\,B=\frac{a^2+c^2-b^2}{2ac},\,C=\frac{a^2+b^2-c^2}{2ab}\,,
\end{equation*} 
and $\dot{\varphi}=d\varphi/dR$.
Using eq.\,\eqref{eq:General Spin Connection}, we obtain
\begin{widetext}
\begin{align*}
    \slashed{\mathcal{D}}_\mathcal{A}=\gamma^4&\left\{\left[-\frac{\partial_r}{f}+iQ^f_{\rm{em}}\mathcal{A}(E_4)-\frac{1}{2}\left(\frac{\dot{a}}{a}+\frac{\dot{b}}{b}+\frac{\dot{c}}{c}\right)\right]I_4\right.\\
    &+\left.\hspace{0cm}\left[i(E_i+iQ^f_{\rm{em}}\mathcal{A}(E_i))\tau_i+\frac{1}{2}\left(\frac{A}{a}+\frac{B}{b}+\frac{C}{c}\right)\right]\begin{pmatrix}
        -\mathbf{1}_2&0\nn\\
        0&\mathbf{1}_2
    \end{pmatrix}\right\}\,.
\end{align*}
\end{widetext}
Introducing 
\begin{equation*}
    \mathcal{D}_i\equiv X_i+iQ^f_{\rm{em}}\mathcal{A}(X_i),~~\mathcal{A}(X_i)=X_i^\mu\mathcal{A}_\mu\,,~~i=1,2,3\,,
\end{equation*}
we get
\begin{equation*}
\slashed{\mathcal{D}}_\mathcal{A}=\begin{pmatrix}
        0&T_{\mathcal{A}}^\dag\\
        T_{\mathcal{A}}&0
    \end{pmatrix}\,,
\end{equation*}
where 
\begin{equation}\label{eq:TA}
    \begin{aligned}
T_\mathcal{A}&=\left[i\frac{\partial_r}{f}+Q^f_{\rm{em}}\mathcal{A}(E_4)+\frac{i}{2}\left(\frac{\dot{a}}{a}+\frac{\dot{b}}{b}+\frac{\dot{c}}{c}\right)\right]\mathbf{1}_2+iB_\mathcal{A}\,,\nn\\
B_\mathcal{A}&=i\left(\frac{\tau_1\mathcal{D}_1}{a}+\frac{\tau_2\mathcal{D}_2}{b}+\frac{\tau_3\mathcal{D}_3}{c}\right)+\frac{1}{2}\left(\frac{A}{a}+\frac{B}{b}+\frac{C}{c}\right)\mathbf{1}_2\,,
    \end{aligned}
\end{equation}
and 
\begin{equation*}
T_\mathcal{A}^\dag=\left[i\frac{\partial_r}{f}+Q^f_{\rm{em}}\mathcal{A}(E_4)+\frac{i}{2}\left(\frac{\dot{a}}{a}+\frac{\dot{b}}{b}+\frac{\dot{c}}{c}\right)\right]\mathbf{1}_2-iB_\mathcal{A}\,.
\end{equation*}
The operator $T_\mathcal{A}^\dag$ is the adjoint operator of $T_\mathcal{A}$ with respect to the scalar product
\begin{equation*}
    \bra{\phi}\ket{\psi}=\int dV\,\,\phi^\dag\psi=\int e^1\wedge e^2\wedge e^3\wedge e^4\,(\phi^*)^{\rm{T}}\psi\,,
\end{equation*}
where $\phi$ and $\psi$ are $\mathcal{L}_2$ functions. Consider the metric
\begin{equation*}
    ds^2_{\rm{hTN}}\equiv {\rm{V}}ds^2_{\rm{H^3L}}+{\rm{V}}^{-1}\sigma_3^2\,,
\end{equation*}
where 
\begin{equation*}
    ds^2_{\rm{H^3L}}=dr^2+4L^2\sinh^2\left(\frac{r}{2L}\right)(\sigma_1^2+\sigma_2^2)\,,
\end{equation*}
and we have introduced the positive function
\begin{equation*}
    {\rm{V}}\equiv\frac{1}{L}\left(\frac{1}{\beta}+\frac{1}{e^{r/L}-1}\right)=\frac{e^{r/L}+\beta-1}{\beta L(e^{r/L}-1)}\,,
\end{equation*}
where $\beta$ and $L$ are non-negative constant. If we conformally rescale the metric $ds^2_{\rm{hTN}}$ with the conformal factor
\begin{equation*}
    \Lambda\equiv\sqrt{\frac{4L}{\beta}}\frac{1}{(2-\beta)\cosh\left(\frac{r}{2L}\right)+\beta\sinh\left(\frac{r}{2L}\right)}\,,
\end{equation*}
we obtain a 1-parameter family of metrics
\begin{equation}\label{eq:beta metrics}
    ds^2_\beta=\Lambda^2ds^2_{\rm{hTN}}\,.
\end{equation}For $\beta\rightarrow2$ we get, making the change of variables
\begin{equation*}
    r=\rho\, {\rm{arccoth}}\left(\frac{w^2}{\rho^2}\right),~~\rho=2L,
\end{equation*}
the AEH line element
\begin{align}\label{eq:Eguchi-Hanson w}
    ds^2_{\beta=2}&=\frac{dw^2}{1-\left(\rho/w\right)^4}\nn\\
    &+\frac{w^2}{4}\left[\sigma_1^2+\sigma_2^2+\left(1-\left(\frac{\rho}{w}\right)^4\right)\sigma_3^2\right]\,,
\end{align}
while, for $\beta\rightarrow0$ we get the Taub-NUT line element. We are now going to explicitly solve the Dirac zero-mode equation $\slashed{\mathcal{D}}_{\mathcal{A}}\psi=0$ on the manifold defined by the line element in eq.\,\eqref{eq:beta metrics} with the connection of the kind (see eq.\,\eqref{eq:gauge field eguchi hanson})\footnote{We notice that, in the colored case, the gauge connection could be written as $\mathcal{A}=p\mathcal{A}_3^a\sigma_3\tau_a$, where $\tau_a$ is the color $SU(2)$ Pauli matrix.}
\begin{equation}\label{eq:general connection}
    \mathcal{A}=p\mathcal{A}_3\sigma_3=p\frac{\sigma_3}{\beta LV}\xrightarrow[\rm{AEH}]{}\frac{p\rho^2}{w^2}\sigma_3\,,
\end{equation}
where $p$ is a real constant. We furthermore are interested in the case $a=b$. In the presence of the gauge connection given in eq.\,\eqref{eq:general connection}, after some algebra, eq.\,\eqref{eq:TA} becomes
\begin{equation*}
    T_p=i\left[\left(\frac{\partial_r}{f}+\frac{\dot a}{a}+\frac{\dot c}{2c}\right)\mathbf{1}+\frac{\sqrt{V}\lambda}{2\Lambda}P_p\right]\,,
\end{equation*}
where 
\begin{align}\label{eq:Pp definition}
    P_p&=\begin{pmatrix}
        \lambda^{-1}(2iX_3-Q^f_{\rm{em}}\tilde{p})&2iX_-\\
        2iX_+&-\lambda^{-1}(2iX_3-Q^f_{\rm{em}}\tilde{p})
    \end{pmatrix}\nn\\&+\left(\frac{\lambda^2+2}{2\lambda}\right)\mathbf{1}_2\,,
\end{align}
$X_{\pm}=X_1\pm i X_2$ and we have introduced the constants
\begin{equation*}
    \begin{aligned}
    \lambda^{-1}&=2LV\sinh{\left(\frac{r}{2L}\right)},\\
    \tilde{p}&=\frac{p}{\beta LV}\,.
    \end{aligned}
\end{equation*}
Thus, in order to compute the eigenvectors of $\slashed{\mathcal{D}}_{\mathcal{A}}\psi=0$, we have to compute the eigenvectors of $P_p$\,\cite{C_N_Pope_1981}. The Laplace operator
\begin{equation*}
    \Delta_{S^3}=-X_1^2-X_2^2-X_3^2\,,
\end{equation*}
commutes with $iX_3$, thus we can consider the basis of simultaneous eigenvectors
\begin{equation*}
    \left\{\ket{j,m,m'}|\,2j\in \mathbb{Z}, ~~2m,2m'\in\mathbb{Z},~~|m|\leq j,|m'|\leq j\,\right\}\,,
\end{equation*}
\begin{equation*}
    \begin{aligned}
        \Delta_{S^3}\ket{j,m,m'}&=j(j+1)\ket{j,m,m'}\,,\\
        iX_3\ket{j,m,m'}&=m\ket{j,m,m'}\,.
    \end{aligned}
\end{equation*}
$iX_3$ and $iX_{\pm}$ satisty the angular momentum algebra, which means
\begin{equation*}
    \begin{aligned}
        iX_+\ket{j,m,m'}&=\sqrt{(j-m)(j+m+1)}\ket{j,m+1,m'}\,,\\
        iX_-\ket{j,m,m'}&=\sqrt{(j+m)(j-m+1)}\ket{j,m-1,m'}\,.
    \end{aligned}
\end{equation*}
Given the expression for $P_p$ it is quite natural to male the ansatz for its eigenvector to be of the form
\begin{equation}\label{eq:Eigenvector ansatz}
    \begin{pmatrix}
        \alpha\ket{j,m,m'}\\
        \beta\ket{j,m+1,m'}
    \end{pmatrix}\,,
\end{equation}
for some real constants $\alpha$, $\beta$ to be determined. Substituting eq.\,\eqref{eq:Eigenvector ansatz} in the eigenvalue equation for $B_p$ one obtains 
\begin{align*}
    \frac{\alpha}{\beta}&=\frac{1}{2\lambda\sqrt{(j-m)(j+m+1)}}\\
    &\times\Bigg[1+2m-Q^f_{\rm{em}}\tilde{p}\\&\pm\sqrt{(1+2m-Q^f_{\rm{em}}\tilde{p})^2+4\lambda^2(j-m)(j+m+1)}\Bigg]\,,
\end{align*}
and we have the eigenvalues, with multiplicity $2j+1$ ($m'$ is a spectator at this level)
\begin{equation*}
    \frac{\lambda}{2}\pm\frac{1}{\lambda}\sqrt{(1+2m-Q^f_{\rm{em}}\tilde{p})^2+4\lambda^2(j-m)(j+m+1)}\,.
\end{equation*}
The above result holds for $-j\leq m\leq j-1$. Indeed, for $m=j$ we have the eigenvector
\begin{equation*}
    \begin{pmatrix}
        \ket{j,j,m'}\\
    0
    \end{pmatrix},
\end{equation*}
with eigenvalue 
\begin{equation*}
    \frac{\lambda}{2}+\frac{1}{\lambda}(2j+1-Q^f_{\rm{em}}\tilde{p}),
\end{equation*}
and multiplicity $2j+1$. For $m=-j-1$ we have the eigenvector
\begin{equation*}
    \begin{pmatrix}
        0\\
    \ket{j,-j,m'}
    \end{pmatrix},
\end{equation*}
with eigenvalue 
\begin{equation*}
    \frac{\lambda}{2}+\frac{1}{\lambda}(2j+1+Q^f_{\rm{em}}\tilde{p}),
\end{equation*}
and multiplicity $2j+1$.\\
We are now ready to solve the Dirac equation for the zero-modes over the backgrounds given by eq.\,\eqref{eq:beta metrics}
\begin{equation*}
    \slashed{\mathcal{D}}_{p}\psi=\slashed{\mathcal{D}}_{p}\begin{pmatrix}
        \Psi\\\Phi
    \end{pmatrix}=0\,,
\end{equation*}
where $\Psi$ and $\Phi$ are 2-component Weyl spinors. Since
\begin{equation*}
    \slashed{\mathcal{D}}_{p}\psi=\begin{pmatrix}
        0&T_p^\dag\\
        T_p&0
    \end{pmatrix}\begin{pmatrix}
        \Psi\\\Phi
    \end{pmatrix}=0\Rightarrow\begin{cases}
        T_p\Psi=0,\\
        T_p^\dag\Phi=0\,.
    \end{cases}
\end{equation*}
It can be shown that $T_p^\dag$ has trivial kernel, hence we can set $\Phi=0$. Using spherical symmetry, the spinor $\Psi$ has the form
\begin{equation*}
    \Psi=\begin{pmatrix}
        K_1\\K_2
    \end{pmatrix}h(r)\ket{j,m,m'}\,,
\end{equation*}
where $K_{1,2}$ are arbitrary constants and $h(r)$ is a radial function. From the form of $P_p$ in eq.\,\eqref{eq:Pp definition} follows that, since for general $K_{1,2}\neq0$,
\begin{equation*}
    P_p\begin{pmatrix}
        K_1\\K_2
    \end{pmatrix}\ket{j,m,m'}\cancel{\propto}\begin{pmatrix}
        K_1\\K_2
    \end{pmatrix}\ket{j,m,m'}\,,
\end{equation*}
the equation $\slashed{\mathcal{D}}_p\Psi=0$ has non-trivial solutions for
\begin{equation*}
\begin{aligned}
    K_2=0\Rightarrow m=j\,,
    \end{aligned}
\end{equation*}
or
\begin{equation*}
\hspace{-0.2cm}
\begin{aligned}
    K_1=0\Rightarrow m=-j\,.
\end{aligned}
\end{equation*}
As a consequence of eq.\,\eqref{eq:beta metrics}, we have
\begin{align*}
    a=b&=\Lambda a_{\rm{hTN}}=2L\Lambda\sinh\left(\frac{r}{2L}\right)\,,\\c&=\Lambda c_{\rm{hTN}}=\frac{\Lambda}{\sqrt{V}}\,,\\f&=\Lambda f_{\rm{hTN}}=\Lambda\sqrt{V}\,,
\end{align*}
and 
\begin{align}\label{eq:explicit form of Tp}
    T_p=&\frac{i}{2L\Lambda\sqrt{V}}\bigg\lbrace\bigg[-2L\partial_r-\frac{LV_{,r}}{2V}-\coth\left(\frac{r}{2L}\right)\nn\\
    &+\frac{3}{2}\left(\frac{(2-\beta)\sinh{\left(\frac{r}{2L}\right)}+\beta\cosh\left(\frac{r}{2L}\right)}{(2-\beta)\cosh\left(\frac{r}{2L}\right)+\beta\sinh{\left(\frac{r}{2L}\right)}}\right)\bigg]\mathbf{1}_2\nn\\
    &+\frac{P_p}{2\sinh{\left(\frac{r}{2L}\right)}}\bigg\rbrace\,.
\end{align}
For $K_2=0$ 
\begin{equation*}
T_p\Psi=0\,,
\end{equation*} 
has the solution
\begin{align}\label{eq:general solution}
    h(r)=&\mathcal{K} \frac{ \left(e^{r/L}-1\right)^j \left(-\beta +e^{r/L}+1\right)^{3/2}}{\sqrt{\beta +e^{r/L}-1}}\times\nn\\
    &\times e^{-\frac{r}{4L}\left(3+4j+\frac{2}{\beta}(2Q^f_{\rm{em}}p-2j-1)\right)}\,,
\end{align}
where $\mathcal{K}$ is a complex integration constant. In the case $K_1=0$, one obtains the solution in eq.\,\eqref{eq:general solution} with $p$ replaced by $-p$. Concentrating on the case $K_2=0$, for large $r$, $h(r)$ has the asymptotic behavior
\begin{equation*}
    h(r)\approx\exp[-\frac{r}{4L}\left(-1+\frac{2}{\beta}(2Q^f_{\rm{em}}p-2j-1)\right)]\,.
\end{equation*}
We now impose this function to be square integrable with respect to the volume element
\begin{align*}
dV&=fabc\,\sigma_1\wedge\sigma_2\wedge\sigma_3\wedge\,dr\nn\\&\approx e^{-r/L}\,\sigma_1\wedge\sigma_2\wedge\sigma_3\wedge\,dr\,,
\end{align*}
therefore, for $h(r)$ to be $\mathcal{L}_2$ we must have 
\begin{equation*}
    2j+1<\frac{\beta}{2}+2Q^f_{\rm{em}}p\,,
\end{equation*}
that is 
\begin{equation*}
    1\leq2j+1<\frac{\beta}{2}+2Q^f_{\rm{em}}p~~\Rightarrow~~2Q^f_{\rm{em}}p>1-\frac{\beta}{2}\,.
\end{equation*}
Analogously, for $K_1=0$
\begin{equation*}
    1\leq2j+1<\frac{\beta}{2}-2Q^f_{\rm{em}}p~~\Rightarrow~~2Q^f_{\rm{em}}p<1-\frac{\beta}{2}\,.
\end{equation*}
Therefore, for fixed $j$ we have (remember that $m'$ is a spectator at this level) $2j+1$ distinct solutions coming from the allowed values for $m'$, giving 
\begin{equation*}
    {\rm{dim}}({\rm{Ker}}(T_p))=\sum_{2j+1=1}^{[2|Q^f_{\rm{em}}p|+\beta/2]}(2j+1)\,,
\end{equation*}
where $[n]$ is the largest integer strictly smaller than $n$ (so that $[3]=2$ etc). In the AEH case, see eq.\,\eqref{eq:monopole quantization}, $Q^f_{\rm{em}}p$ is an integer, therefore the number of zero-modes of the Dirac operator is
\begin{align}
    |n_+-n_-|&=\sum_{2j+1=1}^{[2|Q^f_{\rm{em}}p|+1]}(2j+1)\nn\\&=|Q^f_{\rm{em}}p|\left(2|Q^f_{\rm{em}}p|+1\right)\,,
\end{align}
furthermore the zero-modes take the form
\begin{equation}
    \psi_+^{(j)}(x)=\begin{pmatrix}
        h^{(j)}(r)\ket{j,j,m'}\\
        0\\
        0\\
        0
    \end{pmatrix},~~Q^f_{\rm{em}}p>0\,
\end{equation}
or
\begin{equation}
\psi_-^{(j)}(x)=\begin{pmatrix}
        0\\
        h^{(j)}(r)\ket{j,-j,m'}\\
        0\\
        0
    \end{pmatrix},~~Q^f_{\rm{em}}p<0\,,
\end{equation}
where
\begin{align*}
    h^{(j)}(r)&=c_1 2^{3 j+4} \left(\frac{L^2}{w^2-4L^2}\right)^{j+\frac{3}{2}}\times\nn\\
    &\times \left(\frac{w^2-4L^2}{w^2+4 L^2}\right)^{\frac{1}{2} ( j+ |Q^f_{\rm{em}}p|+1)}\sqrt{\frac{w^2-4L^2}{w^2}}\,,
\end{align*}
that is choosing $c_1=2^{-3j-4}L^{-2j-3}\mathcal{K}$, we obtain (see eq.\,\eqref{eq:Eguchi-Hanson w})
\begin{align*}
h^{(j)}(r)&=\mathcal{K}~\frac{1}{w[w^4-\rho^4]^{(j+1)/2}}\left(\frac{w^2-\rho^2}{w^2+\rho^2}\right)^{\frac{|Q^f_{\rm{em}}p|}{2}}\,,\\r&=\rho\,\text{arccoth}\left(\frac{w^2}{\rho^2}\right)\,,\,\,\rho=2L\,,
\end{align*}
with $\mathcal{K}$ a normalization factor. That is, we have the normalized mode
\begin{align}\label{eq:ZeroModes in Eguchi Hanson}
        h^{(j)}(r)&={2^{j+1}}\sqrt{\frac{2}{\pi}}\left[\frac{\Gamma \left(j+|Q^f_{\rm{em}}p|+1\right)}{\Gamma (2 j+1) \Gamma \left(|Q^f_{\rm{em}}p|-j\right)}\right]^{1/2}\times\nn\\
        &\times\frac{\rho^{2j+1}}{w[w^4-\rho^4]^{(j+1)/2}}\left(\frac{w^2-\rho^2}{w^2+\rho^2}\right)^{\frac{|Q^f_{\rm{em}}p|}{2}}\,,
\end{align}
where
\begin{equation}
    r=\rho\,\text{arccoth}\left(\frac{w^2}{\rho^2}\right)\,,\,\,\rho=2L\,,
\end{equation}
and 
\begin{equation}
j=0,\dots,\,\frac{2|p\,Q_{\rm{em}}^f|-1}{2}\,.
\end{equation}
Analogously to SU(N) gauge group, the index $\slashed{\mathcal{D}}_{\mathcal{A}}$, could also be computed -- instead of computing explicitly and then counting the zero modes -- in an integral form using the Atiyah-Patodi-Singer (APS) index theorem\,\cite{Atiyah_Patodi_Singer_1975, EGUCHI1980213, Franchetti_2018}.
\section{Closing fermion loops with Yukawa interactions in the AEH manifold}\label{Closing fermion loops with Yukawa}
Having identified the 't Hooft vertex, we have to tie up the fermion zero-modes using all the interactions at disposal. For instance, allowing for a Yukawa-type interaction, we have to evaluate integrals of the kind
\begin{align}\label{eq:masterIntegral}
    \mathcal{I}&=\int d^4x_1 \sqrt{g}\,d^4x_2\sqrt{g}\times\nn\\
    &\times[h^{(j)}(x_1)]^2\mathcal{G}_F(x_1,x_2)[h^{(j')}(x_2)]^2\,,
\end{align}
where $\mathcal{G}_F(x_1,x_2)$ is the Euclidean scalar propagator in the AEH background, computed solving
\begin{align*}
   \left(g^{\mu\nu}_{\rm{EH}}\nabla_\mu\nabla_\nu+m_s^2\right)\mathcal{G}_F(x,x')=-4\pi\,\sqrt{g_{\rm{EH}}}\delta^{(4)}(x-x')\,.
\end{align*}
In principle, to do computations with $O(1)$ accuracy, we should compute numerically the integral in eq.\,\eqref{eq:masterIntegral}; however we will discuss a counting rule for $\pi^2$ factors appearing in instanton diagrams. Indeed, as in the Yang-Mills case, one might think at first glance that instanton amplitudes are highly suppressed by loop factors, nonetheless the meaning of \textit{loop} is not trivial in instanton calculus. Similarly to the flat spacetime case, from the integration over vertices we get
\begin{align}
    \int\,d^4x\sqrt{g}\propto \frac{8}{\pi^2}\int_{\rho}^\infty \,dr\,r^3\,,
\end{align}
in addition to the $1/\pi^2$ provided by momentum integration. Moreover, from the explicit expression of the zero modes on EH, see eq.\,\eqref{eq:ZeroModes in Eguchi Hanson}, the zero-mode wavefunctions do not contain $\pi$ factors other the ones coming from the normalization. Therefore, for $Q_{\rm{em}}^fp>0$, we estimate 
\begin{align}
    \mathcal{I}&\propto\left(\frac{8}{\pi^2}\right)^2\frac{1}{\pi^2}\left(\frac{8}{\pi^2}\right)^2\times\nn\\
    &\times\left[\frac{\Gamma (2 j+1) \Gamma \left(Q_{\rm{em}}^f p-j\right)}{\Gamma \left(j+Q_{\rm{em}}^f p+1\right)}\right]\times\nn\\&\times\left[\frac{\Gamma (2 j'+1) \Gamma \left(Q_{\rm{em}}^f p-j'\right)}{\Gamma \left(j'+Q_{\rm{em}}^f p+1\right)}\right]\nn\\
    &\approx\frac{1}{\pi^2}\mathcal{F}_{Q^f_{\rm{em}}}(j)\mathcal{F}_{Q^f_{\rm{em}}}(j')\,,
\end{align}
where we introduced
\begin{equation}
    \mathcal{F}_{Q^f_{\rm{em}}}(j)=\frac{\Gamma (2 j+1) \Gamma \left(Q_{\rm{em}}^fp-j\right)}{\Gamma \left(j+Q_{\rm{em}}^fp+1\right)}\,.
\end{equation}
\begin{figure}[h!]
\includegraphics[width=0.5\textwidth]{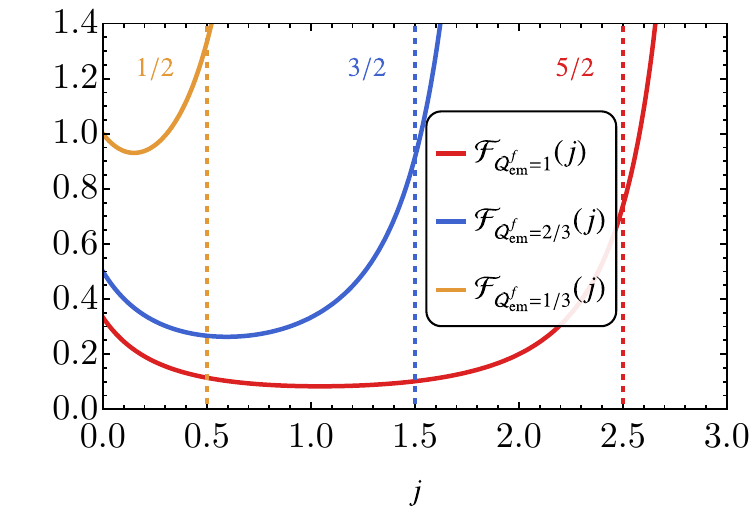}
    \caption{\em Functions $\mathcal{F}_{\mathcal{Q}^f}(j) $ for $\mathcal{Q}_{\rm{em}}^f=1,1/3,2/3$, where from \eqref{eq:ZeroModes in Eguchi Hanson}, $j_{\rm{max}}(\mathcal{Q}_{\rm{em}}=1)=5/2\,,~j_{\rm{max}}(\mathcal{Q}_{\rm{em}}=1/3)=1/2\,,~j_{\rm{max}}(\mathcal{Q}_{\rm{em}}=2/3)=3/2$.}
    \label{fig:FQ}
\end{figure}
Therefore, up to $O(1)$ factors, we can count a loop with a $1/\pi^2$ loop suppression factor (see Figure \ref{fig:FQ}).
\section{The 't Hooft operator}\label{Appendix 't Hooft}
In this Appendix, for completeness, we will review the formalism of the so called 't Hooft vertices in QCD. First, to keep the discussion simple, we consider the one flavor case, $N_f=1$, the Dirac fermions in the fundamental representation of the gauge group (normalized with color Dynkin index $T(\mathcal{C}_q)=1/2$) and the operator
\begin{equation}
    \mathcal{O}=\bar{q}_Lq_R(x),~~\text{with chirality }~\chi=2.
\end{equation}We compute the Euclidean expectation value 
\begin{equation}
    \langle\mathcal{O}(x)\rangle_A=\mathcal{N}\int[\mathcal{D}q][\mathcal{D}\bar{q}]e^{\int d^4\bar{q}i\slashed{\mathcal{D}}q}\bar{q}_Lq_R(x)
\end{equation}
in the presence of an external gauge field $A_\mu^a$, where $\mathcal{N}$ is a normalization factor and $\mathcal{D}_\mu=\delta_\mu+i\,g_S A_\mu$, $\slashed{\mathcal{D}}=\gamma_{\rm{E}}^\mu\mathcal{D}_\mu$. Let $\phi_n(x)$ be the complete set of eigenfunction of the Hermitian eigenvalue problem
\begin{equation}\label{eq:Eigenvalue basis}
\begin{gathered}
    i\slashed{\mathcal{D}}\phi_n(x)=\lambda_n\phi_n(x),\\
    \int d^4x \phi_m(x)^\dag\phi_n(x)=\delta_{nm},\\
    \sum_n\phi_n(x)\phi^\dag_n(y)=\delta^{(4)}(x-y),
\end{gathered}
\end{equation}
where $D_\mu=\partial_\mu-igA_\mu$, with $A_\mu$ some background gauge field. Expanding in the basis \eqref{eq:Eigenvalue basis},
\begin{equation*}
    q(x)=\sum_n\,a_n\phi_n(x),~~~\bar{q}(x)=\sum_n\,\bar{b}_n\phi^\dag_n(x)\,,
\end{equation*}
where, since $q$ and $\bar{q}$ are in the functional integration two independent Grassmann variables, $a_n$ and $\bar{b}_n$ are two independent sets of anti-commuting variables (not carrying spinor indices, which are carried only by the $\phi_n$'s). Therefore, the path integral measure is
\begin{equation*}
    \mathcal{N}\int[\mathcal{D}q][\mathcal{D}\bar{q}]=\mathcal{N}\int\prod_n da_nd\bar{b}_n\equiv\mathcal{N}\int (da)(d\bar{b}),
\end{equation*}
and the fermionic action is 
\begin{equation*}
    -\int d^4x\bar{q}i\slashed{\mathcal{D}}q=\int d^4x\sum_{n,m}\bar{b}_n a_m\left(\phi_n^\dag i\slashed{\mathcal{D}} \phi_m\right)=\sum_{n}\lambda_n\bar{b}_n a_n,
\end{equation*}
therefore
\begin{equation*}
    e^{-S_F}=\prod_n e^{\lambda_n\bar{b}_n a_n}=\prod_n(1+\lambda_n\bar{b}_n a_n),
\end{equation*}
moreover
\begin{equation*}
    \bar{q}_L q_R=\bar{q}\left(\frac{1+\gamma_5}{2}\right)q=\sum_{n,m}\bar{b}_n a_m\phi_n^\dag  \left(\frac{1+\gamma_5}{2}\right)\phi_m.
\end{equation*}
In terms of $a_n$'s and $\bar{b}_n$'s the expectation value of $\mathcal{O}$ then becomes
\begin{align}
    &\langle\mathcal{O}(x)\rangle_A\nn\\
    &=\mathcal{N}\int (da)(d\bar{b})\prod_n(1+\lambda_n\bar{b}_n a_n)\times\nn\\
&\hspace{1cm}\times\sum_{m,l}\bar{b}_m a_l\phi_m^\dag  \left(\frac{1+\gamma_5}{2}\right)\phi_l.
\end{align}
The Grassmann integrals are performed according to the rules
\begin{equation}
    \int da_n\begin{pmatrix}
        1\\
        ~a_m
    \end{pmatrix}=\int d\bar{b}_n\begin{pmatrix}
        1\\
        ~\bar{b}_m
    \end{pmatrix}=\begin{pmatrix}
        0\\
        ~\delta_{nm}
    \end{pmatrix},
\end{equation}
therefore the only non-vanishing contribution to $\langle\mathcal{O}\rangle_A$ are those in which each $a_n$ and $\bar{b}_n$ appears exactly once, giving
\begin{equation}\label{eq:expectation value of O sum}
    \langle\mathcal{O}(x)\rangle_A\propto\prod_m\lambda_n\sum_n\frac{1}{\lambda_n}\phi_n^\dag  \left(\frac{1+\gamma_5}{2}\right)\phi_n(x).
\end{equation}
We now ask what are the external gauge field configurations $A_\mu$ giving a non-vanishing $\langle\mathcal{O}\rangle_A$. That is, we consider the following cases:
\begin{itemize}
    \item  $i\slashed{\mathcal{D}}$ has no zero modes, \textit{i.e.} there are no \textit{normalizable} solutions of the equation $i\slashed{\mathcal{D}}\psi$, in the background of the external gauge field $A_\mu^a$.\\
    This means that\,\cite{Shifman_2022}, \textit{all }the eigenvalues of $i\slashed{\mathcal{D}}$ are paired
    \begin{equation}
        i\slashed{\mathcal{D}}\phi_n=\lambda_n\phi_n,~~~i\slashed{\mathcal{D}}\gamma_5\phi_n=-\lambda_n\gamma_5\phi_n,
    \end{equation}
    and the sum appearing in \eqref{eq:expectation value of O sum} vanishes, i.e.
    \begin{equation}
        \sum_n\frac{1}{\lambda_n}\phi_n^\dag  \left(\frac{1+\gamma_5}{2}\right)\phi_n(x)=0.
    \end{equation}
    That is
    \begin{equation}
         \langle\mathcal{O}(x)\rangle_A=0\,.
    \end{equation}
        \item $i\slashed{\mathcal{D}}$ has one “$+$" chirality zero mode: $n_+=1$.\\
    Calling $\phi_0$ the corresponding eigenvector, i.e. $i\slashed{\mathcal{D}}\phi_0=\lambda_0\phi_0$, $\lambda_0=0$, \eqref{eq:expectation value of O sum} becomes
    \begin{align}
        &\lambda_0\prod_n{}^{'}\lambda_n\Bigg(\frac{1}{\lambda_0}\phi_0^\dag  \left(\frac{1+\gamma_5}{2}\right)\phi_0(x)\nn\\
        &+\underbrace{\frac{1}{\lambda_1}\phi_1^\dag  \left(\frac{1+\gamma_5}{2}\right)\phi_1(x)+\dots}_{=0}\Bigg)\nn\\
        &=\prod_n{}^{'}\lambda_n\phi_0^\dag  \left(\frac{1+\gamma_5}{2}\right)\phi_0(x)\nn\\
        &=\rm{det}'(i\slashed{\mathcal{D}})\phi_0^\dag\left(\frac{1+\gamma_5}{2}\right)\phi_0(x)\nn\\
        &=\rm{det}'(i\slashed{\mathcal{D}})\phi_0^\dag\phi_0(x),
    \end{align}
where, in $\prod^{'}$, we are considering only non-zero modes and we used the fact that eigenfunctions of $i\slashed{D}$ with zero eigenvalue can always be chosen to be eigenfunctions of $\gamma_5$\,\cite{Shifman_2022}. That is
    \begin{equation}
        \langle\mathcal{O}(x)\rangle_A\propto\rm{det}'(i\slashed{\mathcal{D}})\phi_0^\dag\phi_0(x)\neq0,
    \end{equation}
    where $\rm{det}'$ is the determinant computed excluding zero eigenvalues.
\item $i\slashed{\mathcal{D}}$ has one “$-$" chirality zero mode: $n_-=1$.\\
    Proceeding as in the previous point,
    \begin{align}
        \langle\mathcal{O}(x)\rangle_A&\propto\rm{det}'(i\slashed{\mathcal{D}})\phi_0^\dag\left(\frac{1+\gamma_5}{2}\right)\phi_0(x)\nn\\&=0,
    \end{align}
    which means that anti-instanton configurations give vanishing contribution to $\langle\mathcal{O}\rangle_A$.
    \item $i\slashed{\mathcal{D}}$ has more than one zero-mode
    \begin{equation}
    \slashed{\mathcal{D}}\phi^{(n)}_0=\lambda^{(n)}_0\phi^{(n)}_0, ~~\lambda^{(n)}_0=0\,. 
    \end{equation}
    In this case \eqref{eq:expectation value of O sum} yields
    \begin{align}
        &\langle\mathcal{O}(x)\rangle_A\nn\\
        &\propto\lambda_0^{(1)}\lambda_0^{(2)}\dots\lambda_0^{(n)}\rm{det}'(i\slashed{\mathcal{D}})\times\nn\\
        &\times\Bigg(\frac{1}{\lambda_0^{(1)}}\phi_0^{(1)\dag}\left(\frac{1+\gamma_5}{2}\right)\phi_0^{(1)}(x)\nn\\
        &+\frac{1}{\lambda_0^{(2)}}\phi_0^{(2)\dag}\left(\frac{1+\gamma_5}{2}\right)\phi_0^{(2)}(x)+\dots\Bigg)=0,
    \end{align}
    that is gauge field configurations with $|n|=\vert n_+ - n_-\vert\geq2$ do not contribute to the expectation value of $\mathcal{O}$.
\end{itemize}
From the above analysis we conclude that
\begin{equation}\label{eq:chirality and winding number}
\begin{gathered}
    \langle \bar{q}_L q_R\rangle_A\neq0\,,\\\text{requires}~~n=n_+-n_-=1,
\end{gathered}
\end{equation}
similarly
\begin{equation}
\begin{gathered}
    \langle \bar{q}_R q_L\rangle_A\neq0\,,\\\text{requires}~~n=n_+-n_-=-1,
\end{gathered}
\end{equation}
Therefore,
\begin{align}\label{eq:qLqR instanton background}
    &\langle \bar{q}_L q_R\rangle\nn\\&=\mathcal{N}\sum_n\int[\mathcal{D}A]_n[\mathcal{D}\bar{q}][\mathcal{D}q]\,\times\nn\\
    &\hspace{1cm}\times\exp{\int d ^4x\left(-\frac{1}{4}(F_{\mu\nu}^a)^2+\bar{q}i\slashed{\mathcal{D}}q\right)+in\theta}\bar{q}_L q_R(x)\nn\\
    &=\mathcal{N}\int[\mathcal{D}A]_1[\mathcal{D}\bar{q}][\mathcal{D}q]\,\times\nn\\
    &\hspace{1cm}\times\exp{\int d ^4x\left(-\frac{1}{4}(F_{\mu\nu}^a)^2+\bar{q}i\slashed{\mathcal{D}}q\right)+i\theta}\bar{q}_L q_R(x)\nn\\
    &=\mathcal{N}\int[\mathcal{D}A]_1 \exp{-\frac{1}{4}\int d^4x(F_{\mu\nu}^a)^2}e^{i\theta}\frac{\rm{det}'(i\slashed{\mathcal{D}})}{\det(i\slashed\partial)}\,\times\nn\\
    &\hspace{1cm}\times\phi_0^\dag(x)\phi_0(x),
\end{align}
where we have normalized respect to the case with $A_\mu=0$. Substituting the zero-mode wavefunction
\begin{align}\label{eq:explicit fermionic zero mode}
    \phi_0(x)&=\frac{1}{\pi}\frac{\rho}{[(x-x_0)^2+\rho^2]^{3/2}}\begin{pmatrix}
        0\\
        1
    \end{pmatrix}\varphi\,,\nn\\
    \varphi^{\alpha a}&=\varepsilon^{\alpha a},
\end{align}
 Expanding about the instanton solution and substituting the explicit expression for the zero-mode solution \eqref{eq:explicit fermionic zero mode}, we find
\begin{align}\label{eq:q_Lq_R provvisorio}
    \langle \bar{q}_L q_R\rangle&\propto\int\frac{d\rho}{\rho^5}\int d^4x_0\int d\Omega\,\left(\frac{8\pi^2}{g^2}\right)^{2N_c}\,\times\nn\\
    &\times\exp(-\frac{8\pi^2}{g_0^2}+\frac{11}{3}N_c\log(M\rho))e^{i\theta}\frac{\rm{det}'(i\slashed{\mathcal{D}})}{\det(i\slashed\partial)}\,\times\nn\\&\times\frac{\rho^3\varphi^\dag\varphi}{\pi^2\,[(x-x_0)^2+\rho^2]^3}\,,
\end{align}
where $d\Omega$ is the differential corresponding to the color orientation of the instanton and it is normalized to unity. We notice that the $g^2$ factoring \eqref{eq:q_Lq_R provvisorio} is the bare coupling, which we expect to be renormalized by taking into account two-loop effects. Moreover, in \eqref{eq:q_Lq_R provvisorio} we have in the numerator a factor of $\rho^3$, two of which are carried by $\phi_0^\dag\phi_0$ and an extra $\rho$ is inserted for dimensional reasons (to match the dimentions of the LHS of \eqref{eq:q_Lq_R provvisorio}). However the determinants must be regularized from loop divergences, and\,\cite{pseudoparticle}
\begin{align}
    \frac{\rm{det}'(i\slashed{\mathcal{D}})}{\det(i\slashed\partial)}&\rightarrow\left.\frac{\rm{det}'(i\slashed{\mathcal{D}})}{\det(i\slashed\partial)}\right|_{\rm regulated}\nn\\&=\frac{\rm{det}'(i\slashed{\mathcal{D}})}{i(M\rho)\times\rm{det}'(i\slashed{\mathcal{D}}+iM)}\frac{\rm{det}(i\slashed{\partial})}{\rm{det}(i\slashed{\partial}+iM)}\,,\nn\\
    &=\exp\left(\frac{1}{3}\log(M\rho)\right)\,.
\end{align}
Therefore, we finally obtain
\begin{widetext}
\begin{equation}\label{eq:qLqR}
        \langle \bar{q}_L q_R(x)\rangle\propto\left(\frac{8\pi^2}{g^2}\right)^{2N_c}\int\frac{d\rho}{\rho^5}\int d^4x_0\,\left(\frac{8\pi^2}{g^2(1/\rho)}\right)^{2N_c}\,e^{-\frac{8\pi^2}{g^2(1/\rho)}}e^{i\theta}\frac{\rho^3}{[(x-x_0)^2+\rho^2]^3}\,.
\end{equation}
\end{widetext}
We now make the following observation (see\,\cite{PhysRevLett.37.8,SHIFMAN198046}); for small instanton size, the integrand of \eqref{eq:qLqR} goes like
\begin{equation}
    (x-x_0)^{-6},
\end{equation}
and the Euclidean propagator for massless Dirac fermions in coordinate space reads
\begin{equation}
    S_F(x-x_0)=\frac{\gamma_\mu(x-x_0)_\mu}{2\pi^2(x-x_0)^4}\propto(x-x_0)^{-3},
\end{equation}
therefore, for \textit{small instanton} sizes, the amplitude \eqref{eq:qLqR} has exactly the spacetime structure of the amplitude generated by the effective vertex at $x_0$
\begin{equation}\label{eq:prototipe 't Hooft}
\begin{gathered}
    -C\left(\frac{8\pi^2}{g^2}\right)^{2N_c}\int\frac{d\rho}{\rho^5}\,\rho^3 e^{i\theta}\,e^{-\frac{8\pi^2}{g(1/\rho)}}\bar{q}_Rq_L(x_0)\\
    +\rm{h.c.}\,,
\end{gathered}
\end{equation}
where the minus sign comes from the Fermi statistics and the factor 
\begin{equation}
    (x-x_0)^6,
\end{equation}
is precisely reproduced by the two propagators that connect $x_0$ with $x$ and the integral over the collective coordinate $x_0$ will correspond to the integration in coordinate configuration over the vertex variable; the Hermitian conjugate of \eqref{eq:prototipe 't Hooft} gives the anti-instantonic contribution to $\langle\bar{q}_Rq_L(x)\rangle$. This can be visualized diagrammatically as in \eqref{eq:Instanton vertex}.
\tikzset{every picture/.style={line width=0.75pt}} 
\begin{equation}\label{eq:Instanton vertex}
\raisebox{-10mm}{
\begin{tikzpicture}[x=0.75pt,y=0.75pt,yscale=-1,xscale=1]
\draw [line width=1.5]  [fill={rgb, 255:red, 255; green, 222; blue, 172 }  ,fill opacity=1 ] (200,158) .. controls (200,144.19) and (211.19,133) .. (225,133) .. controls (238.81,133) and (250,144.19) .. (250,158) .. controls (250,171.81) and (238.81,183) .. (225,183) .. controls (211.19,183) and (200,171.81) .. (200,158) -- cycle ;
\draw  [line width=1.2]  (236,135.3) .. controls (257.5,137.55) and (282.5,151) .. (287.5,160) ;
\draw [shift={(267.07,144.69)}, rotate = 204.9] [fill={rgb, 255:red, 0; green, 0; blue, 0 }  ][line width=0.08]  [draw opacity=0] (6.25,-3) -- (0,0) -- (6.25,3) -- cycle    ;
\draw [line width=1.2]   (234.5,181.25) .. controls (254.5,180) and (273.5,172.5) .. (287.5,160) ;
\draw [shift={(258.6,176.55)}, rotate = 339.54] [fill={rgb, 255:red, 0; green, 0; blue, 0 }  ][line width=0.08]  [draw opacity=0] (6.25,-3) -- (0,0) -- (6.25,3) -- cycle    ;
\draw  [fill={rgb, 255:red, 0; green, 0; blue, 0 }  ,fill opacity=1 ] (285.5,160) .. controls (285.5,158.9) and (286.4,158) .. (287.5,158) .. controls (288.6,158) and (289.5,158.9) .. (289.5,160) .. controls (289.5,161.1) and (288.6,162) .. (287.5,162) .. controls (286.4,162) and (285.5,161.1) .. (285.5,160) -- cycle ;
\draw (126,149.07) node [anchor=north west][inner sep=0.75pt]  [font=\small]  {$ \begin{array}{l}
Instanton\\
\end{array}$};
\draw (217,153.07) node [anchor=north west][inner sep=0.75pt]  [font=\footnotesize]  {$x_{0}$};
\draw (292.67,153.07) node [anchor=north west][inner sep=0.75pt]  [font=\footnotesize]  {$x$};
\draw (254.67,118.07) node [anchor=north west][inner sep=0.75pt]  [font=\footnotesize]  {$S_{F}( x-x_{0})$};
\draw (254,188.73) node [anchor=north west][inner sep=0.75pt]  [font=\footnotesize]  {$S_{F}( x-x_{0})$};
\end{tikzpicture}}
\end{equation}
Therefore, to capture the instanton effects, neglecting the instanton size $\rho$, we have to add to the QCD Lagrangian the effective vertex
\begin{align}
     \Delta\mathcal{L}(x)&=-\int d\rho\rho^{-5+3}\,d_{N_c}(\rho)e^{i\theta}\,(\mathcal{K}_{N_c}^{(1)})^{i_1i_2}(\bar{q}_Rq_L(x))_{i_1i_2}\nn\\&+\rm{h.c.}\,,
\end{align}
where the function $d_{N_c}(\rho)$ and the tensor $(\mathcal{K}_{N_c}^{(1)})^{i_1i_2}$ are obtained computing the fermionic correlation function $\langle\bar{q}_L q_R(x)\rangle$ with the effective 't Hooft vertex and matching the result with the instanton background computation in \eqref{eq:qLqR instanton background}. For instance (see\,\cite{SHIFMAN198046})
\begin{equation}
\begin{aligned}
(\mathcal{K}_{N_c}^{(1)})^{i_1i_2}&=2\pi^2\delta^{i_1i_2},~~~N_c=2,\\
(\mathcal{K}_{N_c}^{(1)})^{i_1i_2}&=\frac{4}{3}\pi^2\delta^{i_1i_2},~~N_c=3\,.\\
\end{aligned}
\end{equation}
Performing the analogous analysis for the case of 
N flavors, with quarks in the fundamental representation of the color group, we find that the effect of an instanton at small $\rho$ is captured by the fermion effective operator (the determinant, due to Fermi statistics, goes over flavor indices $s,t$)
\begin{align}\label{eq:'t Hooft operator}
    \Delta\mathcal{L}(x)&=-\int d\rho\,\rho^{-5+3N_f}\,d_{N_c}(\rho)(\mathcal{K}_{N_c}^{(N_f)})^{i_1i_2\dots i_{2N_f}}\,\times\nn\\&\times e^{i\theta}\,\det_{s,t}\left({\bar{q}_R^s(x)q^t_L(x)}\right)_{i_1i_2\dots i_{2N_f}}+\rm{h.c.}\,.
\end{align}  
\bibliography{GravAno.bib}
\end{document}